\RequirePackage{fixltx2e}
\documentclass[english]{IEEEtran}
\usepackage[T1]{fontenc}
\usepackage[latin9]{inputenc}
\usepackage{multirow}
\usepackage{amsmath}
\usepackage{amsthm}
\usepackage{amssymb}
\usepackage{graphicx}
\usepackage{geometry}
\makeatletter

\providecommand{\tabularnewline}{\\}

\theoremstyle{definition}
\newtheorem{assumption}{Assumption}
\theoremstyle{remark}
\newtheorem{rem}{\protect\remarkname}
\theoremstyle{plain}
\newtheorem{thm}{\protect\theoremname}

\usepackage{cite}
\usepackage{multirow} 

\makeatother

\usepackage{babel}
\providecommand{\remarkname}{Remark}
\providecommand{\theoremname}{Theorem}

\begin{document}
\global\long\def\rr{\mathbb{R}}%

\title{Simultaneous Online System Identification and Control using Composite
Adaptive Lyapunov-Based Deep Neural Networks}
\author{Omkar Sudhir Patil, Emily J. Griffis, Wanjiku A. Makumi, and Warren
E. Dixon\thanks{The authors are with the Department of Mechanical and Aerospace Engineering,
University of Florida, Gainesville FL 32611-6250 USA. Email: \{patilomkarsudhir,
emilygriffis00, makumiw, wdixon\}@ufl.edu.}\thanks{This research is supported in part by Office of Naval Research Grant
N00014-21-1-2481, AFOSR award number FA8651-21-F-1027, and FA9550-19-1-0169.
Any opinions, findings, and conclusions or recommendations expressed
in this material are those of the author(s) and do not necessarily
reflect the views of the sponsoring agency.}}
\maketitle
\begin{abstract}
Although deep neural network (DNN)-based controllers are popularly
used to control uncertain nonlinear dynamic systems, most results
use DNNs that are pretrained offline and the corresponding controller
is implemented post-training. Recent advancements in adaptive control
have developed controllers with Lyapunov-based update laws (i.e.,
control and update laws derived from a Lyapunov-based stability analysis)
for updating the DNN weights online to ensure the system states track
a desired trajectory. However, the update laws are based on the tracking
error, and offer guarantees on only the tracking error convergence,
without providing any guarantees on system identification. This paper
provides the first result on simultaneous online system identification
and trajectory tracking control of nonlinear systems using adaptive
updates for all layers of the DNN. A combined Lyapunov-based stability
analysis is provided, which guarantees that the tracking error, state-derivative
estimation error, and DNN weight estimation errors are uniformly ultimately
bounded. Under the persistence of excitation (PE) condition, the tracking
and weight estimation errors are shown to exponentially converge to
a neighborhood of the origin, where the rate of convergence and the
size of this neighborhood depends on the gains and a factor quantifying
PE, thus achieving system identification and enhanced trajectory tracking
performance. As an outcome of the system identification, the DNN model
can be propagated forward to predict and compensate for the uncertainty
in dynamics under intermittent loss of state feedback. Comparative
simulation results are provided on a two-link manipulator system and
an unmanned underwater vehicle system with intermittent loss of state
feedback, where the developed method yields significant performance
improvement compared to baseline methods.
\end{abstract}

\section{Introduction}

Deep neural network (DNN)-based methods have garnered popularity as
a means for identification and control of uncertain nonlinear dynamic
systems. Traditional DNN-based control techniques involve initial
offline system identification based on datasets gathered from experimental
trials \cite{Shi.Shi.ea2019,Punjani.Abbeel2015,Bansal.Akametalu.ea2016,Li.Qian.ea2017,Zhou.Helwa.ea2017,Abbeel.Coates.ea2010,Karg2020}.
Subsequently, the identified DNN is used as a model to design controllers
using traditional model-based control techniques. However, the weight
estimates of the DNN are fixed and not updated during task execution,
raising questions about the model's reliability and adaptability.
Moreover, it is often implicitly assumed that minimizing a loss function
would result in the DNN identifying the system dynamics. Whether a
system model can be identified depends on whether the system trajectories
generate information sufficient for the model to be identified, which
manifests in terms of the persistence of excitation (PE) condition
on the model \cite{Sastry1989a}. Although the PE condition is well-studied
and understood in the system identification literature for linear
regression models, only a few recent works remark on the PE condition
for DNNs \cite{Nar.Sastry2019,Nar.Sastry2020,Sridhar.Sokolsky.ea2022,Lamperski2022}.
If the model is not identified, the DNN may not generalize its performance
well beyond the explored trajectories. Consequently, the controller
may not accurately compensate for the uncertainty, thus hazarding
the stability of the closed-loop system.

Recent results in \cite{Joshi.Virdi.ea.2020b,Sun.Greene.ea2021,Le.Greene.ea2021,Patil.Le.ea2022,Le.Patil.ea.2022a,Patil.Le.ea2025,Griffis.Patil.ea23_2,Hart.Griffis.ea2024,Muthirayan.Kharhonekar.2023}
offer online weight updates for the DNN-based controllers to achieve
tracking error convergence. The online weight update laws are derived
from a Lyapunov-based stability analysis, and the corresponding controllers
are popularly known as Lyapunov-based (Lb)-DNN controllers. These
results can achieve tracking error convergence regardless of whether
the PE condition is satisfied. However, these results do not guarantee
parameter estimation. The update laws in these results are based only
on tracking error feedback and are primarily meant to achieve tracking
error convergence. To address this problem, incorporating a prediction
error, i.e., a measure of the discrepancy between the actual dynamics
and their DNN-based estimate, into the adaptation law can help with
parameter estimation. It is desirable to estimate the DNN parameters
to achieve system identification in addition to trajectory tracking,
where the identified model can be used to perform new tasks. For example,
the identified model can be used to predict and compensate for the
uncertain dynamics under intermittent loss of feedback \cite{Chen.Bell.ea2019,Pulido2024,Bell.Sun.ea.2023}.
However, the prediction error is difficult or often impossible to
obtain since the dynamics are unknown and the state-derivative is
typically either unavailable or noisy.

The classical result in \cite{Slotine1989} develops adaptive controllers
with a composite adaptation law that incorporates both tracking and
prediction errors for nonlinear systems with linear-in-parameters
(LIP) uncertainties, where a low-pass filter is applied on both sides
of the dynamics to eliminate the unknown state-derivative term. However,
extending the composite adaptation law from \cite{Slotine1989} to
nonlinear-in-parameters (NIP) uncertainties such as DNNs is challenging
because the inner-layer weights are embedded in nonlinear activation
functions in a nested fashion. Thus, when a low-pass filter is applied
to the dynamics, the resultant expression is not separable in terms
of the model parameters, which introduces technical challenges as
detailed in Appendix VIII.2.

\textbf{Main Contributions. }This paper provides the first result
on simultaneous online system identification and trajectory tracking
control of nonlinear systems using online updates for all layers of
the DNN. The development involves a composite adaptation law based
on a new prediction error formulation using a dynamic state-derivative
observer, which is combined with the tracking error to construct a
least squares-based composite adaptation law. To address the challenges
posed by the nested and NIP structure of DNNs, the Jacobian of the
DNN is used in a composite adaptation law. Then, a first-order Taylor
series expansion of the DNN is used in the analysis to express the
prediction error in terms of the parameter estimation error. Since
the adaptation laws are tightly coupled with the observer and system
dynamics, a combined Lyapunov-based stability analysis is performed
which guarantees the tracking, observer, and parameter estimation
errors are uniformly ultimately bounded (UUB). If the PE condition
is satisfied, the tracking and weight estimation errors are shown
to exponentially converge to a neighborhood of the origin. 

Furthermore, the guarantees on estimating the ideal DNN parameters
imply accurate system identification. Thus, the identified DNN model
can generalize beyond the points encountered by the system trajectory.
As a result, the composite adaptive model is suitable for systems
involving intermittent loss of state feedback, where the identified
DNN model can be propagated forward in time to predict the uncertain
dynamics when feedback is lost, under developed sufficient dwell-time
conditions. To demonstrate the performance and efficacy of the developed
method on different systems, comparative simulation results are provided
on two systems: a robot manipulator and an unmanned underwater vehicle
(UUV) with intermittent loss of state feedback. The developed composite
adaptive Lb-DNN controller yields significant performance improvement
when compared to the tracking error-based adaptive Lb-DNN in \cite{Patil.Le.ea2022}
and state-derivative observer-based disturbance rejection controllers
as baseline methods.

\section{Notation and Preliminaries}

The space of essentially bounded Lebesgue measurable functions is
denoted by $\mathcal{L}_{\infty}$. The identity matrix of size $n$
is denoted by $I_{n}$. The pseudo-inverse of full row rank matrix
$A\in\mathbb{R}^{n\times m}$ is denoted by $A^{+}$, where Given
some matrix $A\triangleq\left[a_{i,j}\right]\in\mathbb{R}^{n\times m}$,
where $a_{i,j}$ denotes the element in the $i^{th}$ row and $j^{th}$
column of $A$, the vectorization operator is defined as $\mathrm{vec}(A)\triangleq[a_{1,1},\ldots,a_{n,1},\ldots,a_{1,m},\ldots,a_{n,m}]^{\top}\in\mathbb{R}^{nm}$.
In the following development, we consider a fully-connected deep neural
network (DNN) $\Phi:\mathbb{R}^{L_{\textrm{in}}}\times\mathbb{R}^{p}\to\mathbb{R}^{L_{\textrm{out}}}$
with $k\in\mathbb{Z}_{>0}$ hidden layers, input size $L_{\textrm{in}}\in\mathbb{Z}_{>0}$,
output size $L_{\textrm{out}}\in\mathbb{Z}_{>0}$, and total number
of parameters $p\in\mathcal{\mathbb{Z}}_{>0}$, where the parameters
include weights and bias terms. Let $\sigma\in\mathbb{R}^{L_{\textrm{in}}}$
denote the DNN input and $\theta\in\mathbb{R}^{p}$ denote the concatenated
vector of DNN parameters. Then, a fully-connected feedforward DNN
$\Phi(\sigma,\theta)$ is defined using a recursive relation $\Phi_{j}\in\mathbb{R}^{L_{j+1}}$
given by 
\begin{eqnarray}
\Phi_{j} & \triangleq & \begin{cases}
V_{j}^{\top}\phi_{j}\left(\Phi_{j-1}\right), & j\in\left\{ 1,\ldots,k\right\} ,\\
V_{j}^{\top}\sigma_{a}, & j=0,
\end{cases}\label{eq:phij_dnn}
\end{eqnarray}
where $\Phi(\sigma,\theta)=\Phi_{k}$ , and $\sigma_{a}\triangleq\left[\begin{array}{cc}
\sigma^{\top} & 1\end{array}\right]^{\top}$ denotes the augmented input that accounts for the bias terms, $V_{j}\in\mathbb{R}^{L_{j}\times L_{j+1}}$
denotes the matrix of weights and biases, and $L_{j}\in\mathbb{Z}_{>0}$
denotes the number of neurons in the $j^{\textrm{th}}$ layer for
all $j\in\left\{ 0,\ldots,k\right\} $ with $L_{0}\triangleq L_{\textrm{in}}+1$
and $L_{k+1}\triangleq L_{\textrm{out}}$. The vector of smooth activation
functions is denoted by $\phi_{j}:\mathbb{R}^{L_{j}}\to\mathbb{R}^{L_{j}}$
for all $j\in\left\{ 1,\ldots,k\right\} $. The activation functions
at each layer are represented as $\phi_{j}\triangleq\left[\varsigma_{j,1}\begin{array}{ccc}
\ldots & \varsigma_{j,L_{j}-1} & \mathrm{1}\end{array}\right]^{\top}$, where $\varsigma_{j,i}:\mathbb{R}\to\mathbb{R}$ denotes the activation
function at the $i^{\mathrm{th}}$ node of the $j^{\mathrm{th}}$
layer. For the DNN architecture in (\ref{eq:phij_dnn}), the vector
of DNN weights is defined as $\theta\triangleq\left[\begin{array}{ccc}
\mathrm{vec}(V_{0})^{\top} & \ldots & \mathrm{vec}(V_{k})^{\top}\end{array}\right]^{\top}$ with size $p=\Sigma_{j=0}^{k}L_{j}L_{j+1}$. The Jacobian of the
DNN with respect to the weights is denoted by $\Phi^{\prime}(\sigma,\theta)\triangleq\frac{\partial}{\partial\theta}\Phi(\sigma,\theta)\in\mathbb{R}^{n\times p}$
is represented as 
\[
\Phi^{\prime}(\sigma,\theta)=\left[\begin{array}{cccc}
\frac{\partial\Phi\left(\sigma,\theta\right)}{\partial\mathrm{vec}(V_{0})}, & \frac{\partial\Phi\left(\sigma,\theta\right)}{\partial\mathrm{vec}(V_{1})}, & \ldots, & \frac{\partial\Phi\left(\sigma,\theta\right)}{\partial\mathrm{vec}(V_{k})}\end{array}\right],
\]
where $\frac{\partial\Phi\left(\sigma,\theta\right)}{\partial\mathrm{vec}(V_{j})}\in\mathbb{R}^{n\times L_{j}L_{j+1}}$.
Then, applying the property $\frac{\partial}{\partial\mathrm{vec}(B)}\mathrm{vec}(ABC)=C^{\top}\otimes A$
to the DNN architecture in (\ref{eq:phij_dnn}) yields 
\begin{equation}
\frac{\partial\Phi\left(\sigma,\theta\right)}{\partial\mathrm{vec}(V_{j})}=\left(\overset{\curvearrowleft}{\prod_{l=j+1}^{k}}V_{l}^{\top}\phi_{l}^{\prime}\left(\Phi_{l-1}\right)\right)\left(I_{L_{j+1}}\otimes\varphi_{j}^{\top}\right),\label{eq:Phiprime_j}
\end{equation}
where $\varphi_{j}$ is a shorthand notation defined as $\varphi_{0}\triangleq\sigma_{a}$
and $\varphi_{j}\triangleq\phi_{j}\left(\Phi_{j-1}\right)$ for all
$j\in\{1,\ldots,k\}$, and the Jacobian of the activation function
vector at the $j^{\mathrm{th}}$ layer is denoted by $\phi_{j}^{\prime}:\mathbb{R}^{L_{j}}\to\mathbb{R}^{L_{j}\times L_{j}}$
and is defined as $\phi_{j}^{\prime}(y)\triangleq\frac{\partial}{\partial y}\phi_{j}(y)$.
Specifically, $\phi_{j}^{\prime}$ evaluates as $\phi_{j}^{\prime}=\mathrm{diag}\left(\left[\varsigma_{j,1}^{\prime}\begin{array}{ccc}
\ldots & \varsigma_{j,L_{j}-1}^{\prime} & \mathrm{0}\end{array}\right]^{\top}\right)$, where $\varsigma_{j,i}^{\prime}(\zeta)\triangleq\frac{\partial}{\partial\zeta}\varsigma_{j,i}(\zeta)$,
and $\mathrm{diag}(\cdot)$ represents the diagonalization operation
which returns a diagonal matrix with the elements of its input vector
arranged along the diagonal. . In (\ref{eq:Phiprime_j}), the notation
$\stackrel{\curvearrowleft}{\prod}$ denotes the right-to-left matrix
product operation, i.e., ${\displaystyle \overset{\curvearrowleft}{\prod}_{p=1}^{m}A_{p}=A_{m}\ldots A_{2}A_{1}}$
and ${\displaystyle \overset{\curvearrowleft}{\prod}_{p=a}^{m}A_{p}=I}$
if $a>m$, and $\otimes$ denotes the Kronecker product. To facilitate
the subsequent development and analysis, the following assumption
is made regarding the activation functions of the DNN.
\begin{assumption}
\label{assm:activation bounds} For each $j\in\left\{ 0,\ldots,k\right\} $,
the activation function $\phi_{j}$, its Jacobian $\phi_{j}^{\prime}$,
and Hessian $\phi_{j}^{\prime\prime}\left(y\right)\triangleq\frac{\partial^{2}}{\partial y^{2}}\phi_{j}\left(y\right)$
are bounded as $\left\Vert \phi_{j}\left(y\right)\right\Vert \leq\mathfrak{a}_{1}\left\Vert y\right\Vert +\mathfrak{a}_{0}$,
$\left\Vert \phi_{j}^{\prime}\left(y\right)\right\Vert \leq\mathfrak{b}_{0}$,
and $\left\Vert \phi_{j}^{\prime\prime}\left(y\right)\right\Vert \leq\mathfrak{c}_{0}$,
respectively, where $\mathfrak{a}_{0},\mathfrak{a}_{1},\mathfrak{b}_{0},\mathfrak{c}_{0}\in\mathbb{R}_{>0}$
are known constants.
\end{assumption}
\begin{rem}
\label{rem:activation bounds} Most activation functions used in practice
satisfy Assumption \ref{assm:activation bounds}. Specifically, sigmoidal
activation functions (e.g., logistic function, hyperbolic tangent
etc.) have $\left\Vert \phi_{j}\left(y\right)\right\Vert $, $\left\Vert \phi_{j}^{\prime}\left(y\right)\right\Vert $,
and $\left\Vert \phi_{j}^{\prime\prime}\left(y\right)\right\Vert $
bounded uniformly by constants. Smooth approximations of rectified
linear unit (ReLUs) such as Swish grow linearly, and hence satisfy
the bound $\left\Vert \phi_{j}\left(y\right)\right\Vert \leq\mathfrak{a}_{1}\left\Vert y\right\Vert +\mathfrak{a}_{0}$
of Assumption \ref{assm:activation bounds}. 
\end{rem}
The following section formulates the trajectory tracking control problem
and provides a control design based on the aforementioned DNN architecture.\footnote{A fully-connected DNN is described here for simplicity in the illustration.
The following control and adaptation law development can be generalized
for any network architecture $\Phi$ with a corresponding Jacobian
$\Phi^{\prime}$. The reader is referred to \cite{Patil.Le.ea2025}
and \cite{Griffis.Patil.ea23_2} for extending the subsequent development
to ResNets and LSTMs, respectively. } 

\section{Problem Formulation and Control Design}

Consider the second order nonlinear system 
\begin{eqnarray}
\ddot{x} & = & f(x,\dot{x})+g(x,\dot{x})u,\label{eq: dynamic system}
\end{eqnarray}
where $x,\dot{x}\in\mathbb{R}^{n}$ denote the states with available
measurements, $\ddot{x}\in\rr^{n}$ is the unknown state-derivative,
$f:\mathbb{R}^{n}\times\mathbb{R}^{n}\rightarrow\mathbb{R}^{n}$ denotes
an unknown continuously differentiable drift function, $g:\mathbb{R}^{n}\times\mathbb{R}^{n}\rightarrow\mathbb{R}^{n\times m}$
denotes a known locally Lipschitz control effectiveness matrix, and
$u\in\mathbb{R}^{m}$ denotes the control input. Let the tracking
error $e\in\mathbb{R}^{n}$ be defined as 
\begin{eqnarray}
e & \triangleq & x-x_{d}{(t)},\label{eq: e}
\end{eqnarray}
where $x_{d}:\mathbb{R}_{\ge0}\rightarrow\mathbb{R}^{n}$ denotes
a smooth reference trajectory that is designed to satisfy $\left\Vert x_{d}{(t)}\right\Vert \leq\overline{x_{d}}$,
$\left\Vert \dot{x}_{d}{(t)}\right\Vert \leq\overline{\dot{x}_{d}}$,
and $\left\Vert \ddot{x}_{d}{(t)}\right\Vert \leq\overline{\ddot{x}_{d}}$
where $\overline{x_{d}},\overline{\dot{x}_{d}},\overline{\ddot{x}_{d}}\in\mathbb{R}_{>0}$
are user-selected constants. The control objective is to design a
DNN-based adaptive controller with a composite adaptation law that
achieves tracking and parameter estimation. For ease of exposition,
the notation $X\triangleq\left[\begin{array}{cc}
x^{\top} & \dot{x}^{\top}\end{array}\right]^{\top}\in\mathbb{R}^{2n}$ is introduced. To aid the subsequent development, the following assumptions
are made.
\begin{assumption}
\label{assm:Function g}The function $g$ is full row rank, and its
right pseudoinverse $g^{+}:\mathbb{R}^{n}\times\mathbb{R}^{n}\rightarrow\rr^{m\times n}$
given by $g^{+}(x,\dot{x})\triangleq g(x,\dot{x})^{\top}\left(g(x,\dot{x})g(x,\dot{x})^{\top}\right)^{-1}$
is assumed to be bounded.
\end{assumption}
\begin{assumption}
\label{assm:Function f growth} The drift function $f$ is such that
there exist strictly increasing smooth functions $\varrho_{1},\varrho_{2}:\mathbb{R}_{\geq0}\to\mathbb{R}_{\geq0}$
satisfying $\left\Vert \frac{\partial f}{\partial x}(x,\dot{x})\right\Vert \leq\varrho_{1}\left(\left\Vert X\right\Vert \right)$
and $\left\Vert \frac{\partial f}{\partial\dot{x}}(x,\dot{x})\right\Vert \leq\varrho_{2}\left(\left\Vert X\right\Vert \right)$.
\end{assumption}
Assumption \ref{assm:Function g} implies the system is not underactuated.
Many electromechanical systems satisfy this assumption, e.g., the
robot manipulator and UUV considered in Section \ref{sec:Simulations}
of this paper, Stewart platforms, hexapod robots, etc. The developed
method can be extended on a case-by-case basis to underactuated systems
using standard nonlinear control tools (e.g., backstepping) unique
for such underactuated systems. Because a universal closed-form stabilizing
nonlinear controller cannot be obtained for an arbitrary underactuated,
system even with perfect model knowledge, the derivation has to be
done on a case-by-case basis for each specific underactuated system,
depending on how $g$ is structured. Assumption \ref{assm:Function f growth}
is reasonable for continuously differentiable $f$ because, for most
practical systems, the bounding functions $\varrho_{1},\varrho_{2}$
can be constructed using basis functions such as polynomials or exponentials.
This assumption is made to obtain accuracy guarantees on a high-gain
state-derivative estimator constructed in the subsequent development. 

\subsection{Control Development}

To facilitate the control development, let the auxiliary error $r\in\mathbb{R}^{n}$
be defined as
\begin{eqnarray}
r & \triangleq & \dot{e}+\alpha_{1}e,\label{eq:r}
\end{eqnarray}
where $\alpha_{1}\in\mathbb{R}_{>0}$ denotes a constant control gain.
Taking the time-derivative on both sides of (\ref{eq:r}), and substituting
(\ref{eq: dynamic system})-(\ref{eq:r}) yields 
\begin{eqnarray}
\dot{r} & = & f(x,\dot{x})+g(x,\dot{x})u-\ddot{x}_{d}{(t)}+\alpha_{1}\left(r-\alpha_{1}e\right).\label{eq:rdot 1}
\end{eqnarray}
DNNs are a powerful tool for approximating unstructured uncertainties,
such as $f$, based on their universal function approximation capabilities
on compact sets \cite{Kidger.Lyons2020}. To this end, consider a
compact set $\Omega\subset\mathbb{R}^{n}$ that will be explicitly
defined later in the development. Additionally, let $\Phi:\mathbb{R}^{2n}\times\mathbb{R}^{p}\to\mathbb{R}^{n}$
denote a general DNN architecture, where $p\in\mathbb{Z}_{>0}$ denotes
the total number of DNN parameters. To formulate the DNN-based approximation,
let the loss function $\mathcal{L}:\mathbb{R}^{p}\to\mathbb{R}_{\geq0}$
be defined as
\begin{equation}
\mathcal{L}\left(\theta\right)\triangleq\int_{\Omega}\left(\left\Vert f\left(x,\dot{x}\right)-\Phi\left(X,\theta\right)\right\Vert ^{2}+\varsigma\left\Vert \theta\right\Vert ^{2}\right)d\mu\left(x\right),\label{eq:Loss Function}
\end{equation}
where $\mu$ denotes the Lebesgue measure, $\varsigma\in\mathbb{R}_{>0}$
denotes a regularizing constant, and the term $\varsigma\left\Vert \theta\right\Vert ^{2}$
represents $L_{2}$ regularization (also popularly known as ridge
regression in the machine learning community) \cite[Sec. 7.1.1]{Goodfellow.Bengio.ea2016}.
Additionally, a user-selected compact convex parameter search space
$\Theta\subset\mathbb{R}^{p}$ satisfying $0_{p}\in\Theta$ is considered
with a bound on the search space $\bar{\theta}=\underset{\theta\in\Theta}{\max}\left\Vert \theta\right\Vert $.
The objective is to identify the vector of ideal DNN parameters $\theta^{*}\in\Theta$
defined as
\begin{eqnarray}
\theta^{*} & \triangleq & \underset{\theta\in\Theta}{\arg\min}\,\mathcal{L}\left(\theta\right).\label{eq:theta star}
\end{eqnarray}
Note that $\mathcal{L}(\theta)$ is not computed online; it only
serves as a theoretical construct to define $\theta^{*}$. The subsequently
defined adaptation laws use only instantaneous values of $r$ and
the subsequently defined prediction errors and DNN Jacobian. For the
parameter estimation problem to be well-posed, the solution $\theta^{*}$
to (\ref{eq:theta star}) is considered to be unique. To this end,
the following assumption is made.
\begin{assumption}
\label{assm:Loss Function Strictly Convex}The loss function $\mathcal{L}$
is strictly convex over the set $\Theta$.\footnote{The assumption of local strict convexity allows the subsequent development
to be analyzed from a convex optimization perspective, which otherwise
would be non-convex due to the nested NIP structure of DNNs. We note
that Assumption \ref{assm:Loss Function Strictly Convex} is a sufficient
condition enabled by regularization rather than a generic property
of DNN loss landscapes. Specifically, due to the strict convexity
of the regularizing term $\varsigma\left\Vert \theta\right\Vert ^{2}$
in (\ref{eq:Loss Function}), there exists $\varsigma\in\mathbb{R}_{>0}$
which ensures $\mathcal{L}\left(\theta\right)$ is convex for all
$\theta\in\Theta$. Additionally, the regularizing term has other
advantages such as mitigation of overfitting \cite[Sec. 7.1.1]{Goodfellow.Bengio.ea2016}.
However, selecting very high values of $\varsigma$ can be counterproductive
as it can obscure the contribution of the $\left\Vert f\left(x,\dot{x}\right)-\Phi\left(X,\theta\right)\right\Vert ^{2}$
term to the loss function while also causing underfitting \cite[Sec. 7.1.1]{Goodfellow.Bengio.ea2016};
therefore, there is a tradeoff between selecting low vs. high values
of $\varsigma$. For further information on when the loss is strictly
convex with a substantial contribution from the $\left\Vert f\left(x,\dot{x}\right)-\Phi\left(X,\theta\right)\right\Vert ^{2}$
term, the reader is referred to the identifiability conditions derived
in \cite[Sec. II.A]{Hart.Patil.ea2025}. }
\end{assumption}
\begin{rem}
\label{rem:Universal Function Approximation} Notice that the universal
function approximation property of DNNs was not invoked in the definition
of $\theta^{*}$. The universal function approximation theorem \cite[Theorem 3.1]{Kidger.Lyons2020}
states that the function space of DNNs is dense in the space of continuous
functions $\mathcal{C}\left(\Omega\right)$. As a result, for any
prescribed $\bar{\varepsilon}>0$, there exists a DNN $\Phi$ and
a corresponding parameter $\theta$ such that $\underset{x\in\Omega}{\max}\left\Vert f\left(x,\dot{x}\right)-\Phi\left(X,\theta\right)\right\Vert <\bar{\varepsilon}$,
and therefore $\int_{\Omega}\left\Vert f\left(x,\dot{x}\right)-\Phi\left(X,\theta\right)\right\Vert ^{2}d\mu\left(x\right)<\bar{\varepsilon}^{2}\mu\left(\Omega\right)$.
However, the universal function approximation theorem provides no
guidance on determining the required network architecture or parameter
search space $\Theta$ for an arbitrarily prescribed accuracy $\bar{\varepsilon}$.
Therefore, we allow $\Theta$ to be arbitrarily constructed in the
analysis, at the loss of guarantees on the approximation accuracy.
Although the accuracy $\bar{\varepsilon}$ which bounds $\underset{x\in\Omega}{\max}\left\Vert f\left(x,\dot{x}\right)-\Phi\left(X,\theta\right)\right\Vert $
might no longer be arbitrary in this case, it would still be finite
due to the continuity of $f$ and $\Phi$, where minimizing the loss
in (\ref{eq:Loss Function}) would achieve the best regularized approximation
of $f$. This formulation provides a well-posed optimization problem
with guaranteed existence and uniqueness of the solution without requiring
knowledge of how to construct $\Theta$ for arbitrary accuracy specifications.
\end{rem}
Due to Remark \ref{rem:Universal Function Approximation}, the drift
function can be modeled as
\begin{eqnarray}
f(x,\dot{x}) & = & \Phi(X,\theta^{*})+\varepsilon(X),\label{eq:UFAP}
\end{eqnarray}
where $\varepsilon:\mathbb{R}^{2n}\rightarrow\mathbb{R}^{n}$ denotes
an unknown function reconstruction error that can be bounded as $\max_{X\in\Omega}\left\Vert \varepsilon(X)\right\Vert \leq\overline{\varepsilon}$.
Based on (\ref{eq:rdot 1}) and the subsequent analysis, the control
input is designed as 
\begin{equation}
u=g^{+}(x,\dot{x})(\ddot{x}_{d}(t)-(\alpha_{1}+k_{r})r+(\alpha_{1}^{2}-1)e-\Phi(X,\hat{\theta})),\label{eq:control input}
\end{equation}
where $k_{r}\in\mathbb{R}_{>0}$ denotes a constant control gain,
and $\hat{\theta}\in\rr^{p}$ denotes the adaptive estimate of the
DNN weights $\theta^{*}$ that is developed using subsequently designed
adaptation laws. Substituting (\ref{eq:UFAP}) and (\ref{eq:control input})
into (\ref{eq:rdot 1}) yields 
\begin{eqnarray}
\dot{r} & = & \Phi(X,\theta^{*})-\Phi(X,\hat{\theta})+\varepsilon(X)-e-k_{r}r.\label{eq:rdot 2}
\end{eqnarray}

\subsection{Composite Adaptation Law}

The classical result in \cite{Slotine1989} develops a composite adaptation
law using tracking and prediction errors for robot manipulators that
involve linearly parameterized uncertainties in the absence of exogenous
disturbances. However, for NIP uncertainties such as DNNs, the traditional
development of the prediction error is not applicable and a new approach
is required. Hence, an innovation of this paper is a new prediction
error formulation based on a dynamic state-derivative estimator that
provides an estimate of the ground truth value of the drift $f$ (c.f.,
\cite{Kamalapurkar.Reish.ea2017}). The dynamic state-derivative observer
is designed as 
\begin{eqnarray}
\dot{\hat{r}} & = & g(x,\dot{x})u-\ddot{x}_{d}{(t)}+\alpha_{1}\left(r-\alpha_{1}e\right)+\hat{f}+\alpha_{2}\tilde{r},\nonumber \\
\dot{\hat{f}} & = & k_{f}\left(\dot{\tilde{r}}+\alpha_{2}\tilde{r}\right)+\tilde{r},\label{eq:observer_2}
\end{eqnarray}
where $\hat{r},\hat{f}\in\rr^{n}$ denote the observer estimates of
$r$ and $f$, respectively, $\tilde{r},\tilde{f}\in\rr^{n}$ denote
the observer errors $\tilde{r}\triangleq r-\hat{r}$ and $\tilde{f}\triangleq f(x,\dot{x})-\hat{f}$,
respectively, and $\alpha_{2},k_{f}\in\rr_{>0}$ denote constant observer
gains. As is typical of observer designs, observer error $\tilde{r}$
is known because $r$ and $\hat{r}$ are known, and is used as feedback
to the observer in (9) to estimate $f$. Since $\dot{\tilde{r}}$
is unknown, (\ref{eq:observer_2}) can be implemented by integrating
both sides and using the relation $\int_{t_{0}}^{t}\dot{\tilde{r}}\left(\tau\right)d\tau=\tilde{r}(t)-\tilde{r}(t_{0})$
to obtain $\hat{f}(t)=\hat{f}(t_{0})+k_{f}\tilde{r}(t)-k_{f}\tilde{r}(t_{0})+\int_{t_{0}}^{t}(k_{f}\alpha_{2}+1)\tilde{r}(\tau)d\tau$,
where $t_{0}$ denotes the initial time. Note that although $\hat{f}$
generated by the state-derivative estimator can also be used to compensate
for $f$, such an approach results in a robust high-gain design which
does not achieve the system identification objective and can cause
large overshoots in the control input.

Taking the time-derivative of $\tilde{r}$ and $\tilde{f}$ and substituting
their definitions along with (\ref{eq:rdot 1}) and (\ref{eq:observer_2})
yields 
\begin{equation}
\dot{\tilde{r}}=\tilde{f}-\alpha_{2}\tilde{r},\qquad\dot{\tilde{f}}=\dot{f}-k_{f}\tilde{f}-\tilde{r},\label{eq:r tilde dot}
\end{equation}
where $\dot{f}\triangleq\frac{\partial f}{\partial x}\dot{x}+\frac{\partial f}{\partial\dot{x}}\ddot{x}$,
and $\dot{\tilde{f}}$ is derived after substituting in $\dot{\tilde{r}}$.
Using the dynamic state-derivative estimator, the prediction error
$E\in\rr^{n}$ is designed as 
\begin{eqnarray}
E & \triangleq & \hat{f}-\Phi\left(X,\hat{\theta}\right).\label{eq:pred error}
\end{eqnarray}
Then, the composite least squares adaptation law is designed as 
\begin{equation}
\dot{\hat{\theta}}=\mathrm{proj}\left(-k_{\hat{\theta}}\Gamma{(t)}\hat{\theta}+\Gamma{(t)}\Phi^{\prime\top}\left(X,\hat{\theta}\right)\left(r+\alpha_{3}E\right)\right),\label{eq:adapt law}
\end{equation}
where $\mathrm{proj}\left(\cdot\right)$ denotes a continuous projection
operator (cf. \cite[Appendix E]{Krstic.Kanellakopoulos.ea1995}) which
ensures $\hat{\theta}(t)\in\mathcal{B}_{\bar{\theta}}\triangleq\{\theta\in\rr^{p}:\left\Vert \theta\right\Vert \leq\bar{\theta}\}$
for all $t\in\rr_{\geq0}$, $\alpha_{3},k_{\hat{\theta}}\in\mathbb{R}_{>0}$
denote constant gains, $\Phi^{\prime}\left(X,\hat{\theta}\right)\in\rr^{n\times p}$
denotes the Jacobian $\Phi^{\prime}\left(X,\hat{\theta}\right)\triangleq\frac{\partial\Phi\left(X,\hat{\theta}\right)}{\partial\hat{\theta}}$.
Similar development could also be used to derive the Jacobian for
other DNN architectures. The term $\Gamma{:\mathbb{R}_{\geq t_{0}}\to}\rr^{p\times p}$
denotes a positive-definite (PD) time-varying least squares adaptation
gain matrix that is a solution to \cite[Eqns. (16) and (17)]{Slotine1989}
{\small
\begin{equation}
\frac{d}{dt}\Gamma^{-1}=\begin{cases}
-\beta(t)\Gamma^{-1} & \lambda_{\Gamma,\min}<\lambda_{\min}\left(\Gamma\right)\\
+\Phi^{\prime\top}\left(X,\hat{\theta}\right)\Phi^{\prime}\left(X,\hat{\theta}\right), & \mathrm{and}\,\lambda_{\max}\left(\Gamma\right)<\lambda_{\Gamma,\max}\\
0_{p\times p}, & \mathrm{otherwise},
\end{cases}\label{eq:Gamma update}
\end{equation}
}where $\lambda_{\Gamma,\min},\lambda_{\Gamma,\max}\in\mathbb{R}_{>0}$
are user-prescribed lower and upper bounds on the minimum and maximum
eigenvalues of $\Gamma$, respectively. The term $\beta:\rr_{\geq0}\to\rr_{\geq0}$
represents a bounded-gain time-varying forgetting factor designed
as 
\begin{equation}
\beta(t)\triangleq\beta_{0}\left(1-\frac{\left\Vert \Gamma(t)\right\Vert }{\varkappa_{0}}\right),\label{eq:beta}
\end{equation}
where $\beta_{0}\in\rr_{>0}$ are user-defined constants denoting
the maximum forgetting rate and the bound on $\lambda_{\max}\left\{ \Gamma(t)\right\} $,
respectively. The adaptation gain in (\ref{eq:Gamma update}) is initialized
to be PD such that $\lambda_{\Gamma,\min}<\left\Vert \Gamma(t_{0})\right\Vert <\varkappa_{0}\leq\lambda_{\Gamma,\max}$,
and it can be shown that $\Gamma{(t)}$ remains PD with $\lambda_{\Gamma,\min}<\left\Vert \Gamma(t)\right\Vert <\varkappa_{0}\leq\lambda_{\Gamma,\max}$
for all $t\in\rr_{\geq t_{0}}$ \cite{Slotine1989}. The term $\beta(t)$
can be lower bounded as $\beta\geq\beta_{1}$, where $\beta_{1}\in\rr_{\geq0}$
is a constant which satisfies the properties stated in the following
remark. 
\begin{rem}
\label{rem:PE property} Consider the case when $\Phi^{\prime}\left(X,\hat{\theta}\right)$
satisfies the uniform PE condition, i.e., there exists constants $\varphi_{1},\varphi_{2}\in\rr_{>0}$
for all $\underline{t}\in\rr_{\geq t_{0}}$ and some $T\in\rr_{>0}$
such that $\varphi_{1}I_{p}\leq\int_{\underline{t}}^{\underline{t}+T}\Phi^{\prime\top}\left(X(\tau),\hat{\theta}(\tau)\right)\Phi^{\prime}\left(X(\tau),\hat{\theta}(\tau)\right)d\tau\leq\varphi_{2}I_{p}$
for all $\left(X(t_{0}),\hat{\theta}(t_{0})\right)\in\mathbb{R}^{n}\times\mathbb{R}^{p}$.
In this case, it can be shown that $\beta_{1}>0$ \cite[Sec. 4.2]{Slotine1989}. 
\end{rem}
The following section shows the stability analysis for the developed
DNN-based composite adaptive control method over the time-interval
$[t_{0},\infty)\subseteq\rr_{\geq0}$.

\section{\label{sec:Stability-Analysis}Stability Analysis}

DNNs are nonlinear with respect to the weights. Designing adaptive
controllers and performing stability analyses for systems that are
NIP has historically been a challenging task. A method to address
the NIP structure of the uncertainty, especially for DNNs, is to use
a first-order Taylor series approximation \cite{Patil.Le.ea2022}.
Let $\tilde{\theta}\triangleq\theta^{*}-\hat{\theta}\in\rr^{p}$ denote
the parameter estimation error. Applying Taylor's first-order theorem
\cite[Theorem 5.15]{Rudin1976} yields
\begin{equation}
\Phi(X,\theta^{*})-\Phi\left(X,\hat{\theta}\right)=\Phi^{\prime}\left(X,\hat{\theta}\right)\tilde{\theta}+R\left(X,\tilde{\theta}\right),\label{eq:Taylor_approx}
\end{equation}
where $R\left(X,\tilde{\theta}\right)\in\mathbb{R}^{n}$ denotes the
Lagrange remainder. Substituting (\ref{eq:Taylor_approx}) into (\ref{eq:rdot 2})
yields the closed-loop error system 
\begin{eqnarray}
\dot{r} & = & \Phi^{\prime}\left(X,\hat{\theta}\right)\tilde{\theta}+\Delta-e-k_{r}r,\label{eq:rdot 3}
\end{eqnarray}
where $\Delta\in\rr^{n}$ is defined as $\Delta\triangleq R\left(X,\tilde{\theta}\right)+\varepsilon(X)$.
To facilitate the subsequent analysis, the prediction error $E$ in
(\ref{eq:pred error}) can be rewritten by adding and subtracting
$f$, substituting in (\ref{eq:UFAP}) and (\ref{eq:Taylor_approx}),
and using the relation $\hat{f}=f-\tilde{f}$, which yields 
\begin{equation}
E=\Phi^{\prime}\left(X,\hat{\theta}\right)\tilde{\theta}-\tilde{f}+\Delta.\label{eq:pred error analytical}
\end{equation}
Taking the time-derivative of $\tilde{\theta}$, substituting (\ref{eq:adapt law}),
and then applying (\ref{eq:pred error analytical}) and the relation
$\hat{\theta}=\theta^{*}-\tilde{\theta}$ yields the parameter estimation
error dynamics 
\begin{eqnarray}
\dot{\tilde{\theta}} & = & -\mathrm{proj}\left(\begin{array}{c}
\Gamma{(t)}\left(k_{\hat{\theta}}+\alpha_{3}\Phi^{\prime\top}\left(X,\hat{\theta}\right)\Phi^{\prime}\left(X,\hat{\theta}\right)\right)\tilde{\theta}\end{array}\right.\nonumber \\
 &  & +\Gamma{(t)}\Phi^{\prime\top}\left(X,\hat{\theta}\right)r-\alpha_{3}\Gamma{(t)}\Phi^{\prime\top}\left(X,\hat{\theta}\right)\tilde{f}\nonumber \\
 &  & \left.+\alpha_{3}\Gamma{(t)}\Phi^{\prime\top}\left(X,\hat{\theta}\right)\Delta-k_{\hat{\theta}}\Gamma{(t)}\theta^{*}\right).\label{eq:theta tilde dot}
\end{eqnarray}

Let $z\triangleq\left[\begin{array}{ccccc}
e^{\top} & r^{\top} & \tilde{r}^{\top} & \tilde{f}^{\top} & \tilde{\theta}^{\top}\end{array}\right]^{\top}\in\rr^{4n+p}$ denote the concatenated state. It follows from using (\ref{eq: e}),
(\ref{eq:r}), and the fact that $\left\Vert x_{d}(t)\right\Vert \leq\overline{x_{d}}$
and $\left\Vert \dot{x}_{d}(t)\right\Vert \leq\overline{\dot{x}_{d}}$
for all $t\in\mathbb{R}_{\geq0}$, that the state $X$ can be bounded
as
\begin{equation}
\left\Vert X\right\Vert \leq\left(\alpha_{1}+2\right)\left\Vert z\right\Vert +\overline{x_{d}}+\overline{\dot{x}_{d}}.\label{eq:X bound}
\end{equation}
By Taylor's theorem \cite[Theorem 5.15]{Rudin1976}, it can be shown
that if $\bar{H}\in\mathbb{R}_{>0}$ is a bound on the DNN Hessian,
$\left\Vert \frac{\partial^{2}\Phi(X,\theta)}{\partial\theta}\right\Vert \leq\bar{H}$
for all $X\in\Omega$ and $\theta\in\Theta$, then the Lagrange remainder
term can be bounded as $\left\Vert R\left(X,\tilde{\theta}\right)\right\Vert \leq\frac{\bar{H}}{2}\left\Vert \tilde{\theta}\right\Vert ^{2}$.
By examining the structure of the DNN and its Jacobian in (\ref{eq:phij_dnn})
and (\ref{eq:Phiprime_j}), it can be shown under Assumption \ref{assm:activation bounds}
and boundedness of the parameter set $\Theta$ that there exists\footnote{For details on the explicit computation of such a polynomial, the
reader is referred to \cite[Theorem 1]{Patil.Fallin.ea2025}.} a polynomial function $\rho_{0}:\mathbb{R}_{\geq0}\to\mathbb{R}_{\geq0}$
of the form $\rho_{0}\left(\left\Vert X\right\Vert \right)=a_{2}\left\Vert X\right\Vert ^{2}+a_{1}\left\Vert X\right\Vert ^{1}+a_{0}$
with constants $a_{2},a_{1},a_{0}\in\mathbb{R}_{>0}$ such that the
DNN Hessian is bounded as $\left\Vert \frac{\partial^{2}\Phi(X,\theta)}{\partial\theta}\right\Vert \leq\rho_{0}\left(\left\Vert X\right\Vert \right)$.
Therefore, the Lagrange remainder term is bounded as $\left\Vert R\left(X,\tilde{\theta}\right)\right\Vert \leq\rho_{0}\left(\left\Vert X\right\Vert \right)\left\Vert \tilde{\theta}\right\Vert ^{2}$.
Because $\rho_{0}$ is a second order polynomial, there exists a second
order polynomial $\rho_{1}:\mathbb{R}_{\geq0}\to\mathbb{R}_{\geq0}$
such that $\rho_{1}\left(\left\Vert z\right\Vert \right)=\rho_{0}\left(\left(\alpha_{1}+2\right)\left\Vert z\right\Vert +\overline{x_{d}}+\overline{\dot{x}_{d}}\right)$.
Therefore, taking the norm of $\Delta$ and using triangle inequality
and the bounds $\max_{X\in\Omega}\left\Vert \varepsilon(X)\right\Vert \leq\overline{\varepsilon}$
from Remark \ref{rem:Universal Function Approximation} yields 
\begin{eqnarray}
\left\Vert \Delta\left(X,\tilde{\theta}\right)\right\Vert  & \leq & \rho_{1}\left(\left\Vert z\right\Vert \right)\left\Vert \tilde{\theta}\right\Vert ^{2}+\bar{\varepsilon}\nonumber \\
 & \leq & \rho_{1}\left(\left\Vert z\right\Vert \right)\left\Vert z\right\Vert ^{2}+\bar{\varepsilon}.\label{eq:Delta bound}
\end{eqnarray}
Since the universal function approximation property of the DNN stated
in (\ref{eq:UFAP}) holds only on the compact domain $\Omega$, the
subsequent stability analysis requires ensuring $X(t)\in\Omega$ for
all $t\in[t_{0},\infty)$. Due to (\ref{eq:X bound}), $X$ can be
restricted to $\Omega$ by obtaining a stability result which constrains
$z$ in an appropriate compact domain. To this end, the compact domain
in which $z$ is supposed to lie,
\begin{equation}
\mathcal{D}\triangleq\left\{ \iota\in\mathbb{R}^{2n+p}:\left\Vert \iota\right\Vert \leq\chi\right\} ,\label{eq:mathcal D}
\end{equation}
is constructed, where $\mathbb{\chi}\in\mathbb{R}_{>0}$ is a user-selected
constant explicitly defined later in the analysis. Then using (\ref{eq:X bound}),
it follows that if $z\in\mathcal{D}$, then $X\in\Omega$, where $\Omega$
can now explicitly be constructed as
\[
\Omega\triangleq\left\{ \iota\in\mathbb{R}^{2n}:\left\Vert \iota\right\Vert \leq\left(\alpha_{1}+2\right)\chi+\overline{x_{d}}+\overline{\dot{x}_{d}}\right\} .
\]

To facilitate the stability analysis, let $V:\rr^{4n+p}\to\rr_{\geq0}$
be the candidate Lyapunov function defined as 
\begin{equation}
V(z)=\frac{1}{2}e^{\top}e+\frac{1}{2}r^{\top}r+\frac{1}{2}\tilde{r}^{\top}\tilde{r}+\frac{1}{2}\tilde{f}^{\top}\tilde{f}+\frac{1}{2}\tilde{\theta}^{\top}\Gamma^{-1}{(t)}\tilde{\theta},\label{eq:Lyap}
\end{equation}
which satisfies the inequality 
\begin{equation}
\lambda_{1}\left\Vert z\right\Vert ^{2}\leq V(z)\leq\lambda_{2}\left\Vert z\right\Vert ^{2},\label{eq:Lyap bounds}
\end{equation}
where $\lambda_{1}\triangleq\min\{\frac{1}{2},\frac{1}{2\lambda_{\Gamma,\max}}\}{\in\mathbb{R}_{>0}}$
and $\lambda_{2}\triangleq\max\{\frac{1}{2},\frac{1}{2\lambda_{\Gamma,\min}}\}{\in\mathbb{R}_{>0}}$.
Taking the time-derivative of $V(z)$, substituting in (\ref{eq:r}),
(\ref{eq:r tilde dot}), (\ref{eq:Gamma update}), (\ref{eq:rdot 3}),
and (\ref{eq:theta tilde dot}), applying the property of projection
operators $-\tilde{\theta}^{\top}\Gamma^{-1}{(t)}\mathrm{proj}(\mu)\leq-\tilde{\theta}^{\top}\Gamma^{-1}{(t)}\mu$
\cite[Lemma E.1.IV]{Krstic.Kanellakopoulos.ea1995}, and canceling
coupling terms yields 
\begin{align}
\dot{V} & \leq-\alpha_{1}\left\Vert e\right\Vert ^{2}-k_{r}\left\Vert r\right\Vert ^{2}-\alpha_{2}\left\Vert \tilde{r}\right\Vert ^{2}-k_{f}\left\Vert \tilde{f}\right\Vert ^{2}\nonumber \\
 & -\left(k_{\hat{\theta}}+\frac{\beta(t)}{2\lambda_{\Gamma,\max}}\right)\left\Vert \tilde{\theta}\right\Vert ^{2}+r^{\top}\Delta+\tilde{f}^{\top}\dot{f}\nonumber \\
 & -\left(\alpha_{3}-\frac{1}{2}\right)\tilde{\theta}^{\top}\Phi^{\prime\top}\left(X,\hat{\theta}\right)\Phi^{\prime}\left(X,\hat{\theta}\right)\tilde{\theta}\nonumber \\
 & +\alpha_{3}\tilde{\theta}^{\top}\Phi^{\prime\top}\left(X,\hat{\theta}\right)\left(\tilde{f}-\Delta\right)+k_{\hat{\theta}}\tilde{\theta}^{\top}\theta^{*}.\label{eq:Vdot 1}
\end{align}
Because $\bar{\theta}=\underset{\theta\in\Theta}{\max}\left\Vert \theta\right\Vert $
and $\theta^{*}\triangleq\underset{\theta\in\Theta}{\arg\min}\mathcal{L}\left(\theta\right)$,
it follows that $\left\Vert \theta^{*}\right\Vert \leq\bar{\theta}$
. Furthermore, due to the use of the projection operator, $\left\Vert \hat{\theta}\right\Vert \leq\bar{\theta}$.
Hence, $\left\Vert \tilde{\theta}\right\Vert \leq2\bar{\theta}$.
By \cite[Lemma 2]{Patil.Fallin.ea2025}, there exist constants $\upsilon_{1},\upsilon_{2}\in\mathbb{R}_{\geq0}$
such that $\left\Vert \Phi^{\prime}\left(X,\hat{\theta}\right)\right\Vert \leq\upsilon_{1}\left\Vert X\right\Vert +\upsilon_{2}$.
Additionally, $\left\Vert \dot{f}\right\Vert $ can be bounded as
$\left\Vert \dot{f}\right\Vert \leq\left\Vert \frac{\partial f}{\partial x}\right\Vert \left\Vert \dot{x}\right\Vert +\left\Vert \frac{\partial f}{\partial\dot{x}}\right\Vert \left\Vert \ddot{x}\right\Vert \leq\left\Vert \frac{\partial f}{\partial x}\right\Vert \left(\left\Vert \dot{e}\right\Vert +\left\Vert \dot{x}_{d}\right\Vert \right)+\left\Vert \frac{\partial f}{\partial\dot{x}}\right\Vert \left(\left\Vert \dot{r}\right\Vert +\alpha_{1}\left\Vert \dot{e}\right\Vert +\left\Vert \ddot{x}_{d}\right\Vert \right)$.
Therefore, it can be shown that $\left\Vert \dot{f}\right\Vert \leq\rho_{2}\left(\left\Vert z\right\Vert \right),$
where $\rho_{2}:\mathbb{R}_{\geq0}\to\mathbb{R}_{\geq0}$ is a strictly
increasing function defined as 
\begin{eqnarray*}
\rho_{2}\left(\left\Vert z\right\Vert \right) & \triangleq & \varrho_{3}(\left\Vert z\right\Vert )\left((\alpha_{1}+1)\left\Vert z\right\Vert +\overline{\dot{x}_{d}}\right)\\
 &  & +\varrho_{4}(\left\Vert z\right\Vert )\left(\rho_{1}(\left\Vert z\right\Vert )\left\Vert z\right\Vert ^{2}+\left(\overline{x_{d}}\upsilon_{1}+\overline{\dot{x}_{d}}\upsilon_{1}\right.\right.\\
 &  & \left.+\upsilon_{2}+\alpha_{1}^{2}+\alpha_{1}+k_{r}+1\left\Vert z\right\Vert \right)\\
 &  & \left.+\bar{\varepsilon}+\overline{\ddot{x}_{d}}\right)+(\alpha_{1}+2)\upsilon_{1}\left\Vert z\right\Vert ^{2},
\end{eqnarray*}
and $\varrho_{3},\varrho_{4}:\mathbb{R}_{\geq0}\to\mathbb{R}_{\geq0}$
are strictly increasing functions defined as
\[
\varrho_{3}\left(\left\Vert z\right\Vert \right)\triangleq\varrho_{1}((\alpha_{1}+2)\left\Vert z\right\Vert +\overline{x_{d}}+\overline{\dot{x}_{d}})
\]
and
\begin{align*}
\varrho_{4}\left(\left\Vert z\right\Vert \right) & \triangleq\varrho_{2}((\alpha_{1}+2)\left\Vert z\right\Vert +\overline{x_{d}}+\overline{\dot{x}_{d}}),
\end{align*}
and $\varrho_{1},\varrho_{2}$ are defined in Assumption \ref{assm:Function f growth}.
Therefore, using Young's inequality, the term $\tilde{f}^{\top}\dot{f}$
in (\ref{eq:Vdot 1}) can be bounded as $\tilde{f}^{\top}\dot{f}\leq\frac{k_{f}\left\Vert \tilde{f}\right\Vert ^{2}}{2}+\frac{\rho_{2}^{2}\left(\left\Vert z\right\Vert \right)}{2k_{f}}$.
By first-order Taylor's theorem and the chain rule, $\rho_{2}^{2}\left(\left\Vert z\right\Vert \right)$
can be represented as $\rho_{2}^{2}\left(\left\Vert z\right\Vert \right)=\rho_{2}^{2}\left(0\right)+2\rho_{2}\left(0\right)\rho_{2}^{\prime}\left(0\right)\left\Vert z\right\Vert +\left(2\rho_{2}^{\prime2}\left(\tau\right)+2\rho_{2}\left(\tau\right)\rho_{2}^{\prime\prime}\left(\tau\right)\right)\left\Vert z\right\Vert ^{2}$
for some $\tau\in\left(0,\left\Vert z\right\Vert \right)$, where
$\rho_{2}^{\prime},\rho_{2}^{\prime\prime}$ denote the first and
second derivatives of $\rho_{2}$, respectively. Furthermore, because
$\left\Vert z\right\Vert \leq\frac{\left\Vert z\right\Vert ^{2}}{2}+\frac{1}{2}$,
it follows that $2\rho_{2}\left(0\right)\rho_{2}^{\prime}\left(0\right)\left\Vert z\right\Vert \leq\rho_{2}\left(0\right)\rho_{2}^{\prime}\left(0\right)\left\Vert z\right\Vert ^{2}+\rho_{2}\left(0\right)\rho_{2}^{\prime}\left(0\right)$.
Additionally, consider a strictly increasing function $\rho_{3}:\mathbb{R}_{\geq0}\to\mathbb{R}_{\geq0}$
such that $2\rho_{2}^{\prime2}\left(\tau\right)+2\rho_{2}\left(\tau\right)\rho_{2}^{\prime\prime}\left(\tau\right)+\rho_{2}\left(0\right)\rho_{2}^{\prime}\left(0\right)\leq\rho_{3}\left(\left\Vert z\right\Vert \right)$
for all $\tau\in\left(0,\left\Vert z\right\Vert \right)$, implying
$\rho_{2}^{2}\left(\left\Vert z\right\Vert \right)\leq\rho_{2}^{2}\left(0\right)+\rho_{2}\left(0\right)\rho_{2}^{\prime}\left(0\right)+\rho_{3}\left(\left\Vert z\right\Vert \right)\left\Vert z\right\Vert ^{2}$.
Hence, $\tilde{f}^{\top}\dot{f}\leq\frac{k_{f}\left\Vert \tilde{f}\right\Vert ^{2}}{2}+\frac{\rho_{3}\left(\left\Vert z\right\Vert \right)}{2k_{f}}\left\Vert z\right\Vert ^{2}+\frac{\rho_{2}^{2}\left(0\right)+\rho_{2}\left(0\right)\rho_{2}^{\prime}\left(0\right)}{2k_{f}}$.

In (\ref{eq:Vdot 1}), because $r^{\top}\Delta\leq\rho_{1}\left(\left\Vert z\right\Vert \right)\left\Vert z\right\Vert ^{2}\left\Vert r\right\Vert +\bar{\varepsilon}\left\Vert r\right\Vert $,
it follows by multiplying and dividing with $\frac{k_{r}}{2}$ and
using Young's inequality that $\rho_{1}\left(\left\Vert z\right\Vert \right)\left\Vert z\right\Vert ^{2}\left\Vert r\right\Vert \leq\frac{k_{r}}{4}\left\Vert r\right\Vert ^{2}+\frac{\rho_{1}^{2}\left(\left\Vert z\right\Vert \right)\left\Vert z\right\Vert ^{4}}{k_{r}}$
and $\bar{\varepsilon}\left\Vert r\right\Vert \leq\frac{k_{r}}{4}\left\Vert r\right\Vert ^{2}+\frac{\bar{\varepsilon}^{2}}{k_{r}}$.
Therefore $r^{\top}\Delta\leq\frac{k_{r}}{2}\left\Vert r\right\Vert ^{2}+\frac{\rho_{1}^{2}\left(\left\Vert z\right\Vert \right)\left\Vert z\right\Vert ^{4}}{k_{r}}+\frac{\bar{\varepsilon}^{2}}{k_{r}}$.
Furthermore, because $\tilde{\theta}^{\top}\theta^{*}\leq\left(\frac{\left\Vert \tilde{\theta}\right\Vert }{\sqrt{2}}\right)\sqrt{2}\bar{\theta}\leq\frac{1}{4}\left\Vert \tilde{\theta}\right\Vert ^{2}+\bar{\theta}^{2}$
by Young's inequality, it follows that $k_{\hat{\theta}}\tilde{\theta}^{\top}\theta^{*}\leq\frac{k_{\hat{\theta}}}{4}\left\Vert \tilde{\theta}\right\Vert ^{2}+k_{\hat{\theta}}\bar{\theta}^{2}$.
As for the term $\alpha_{3}\tilde{\theta}^{\top}\Phi^{\prime\top}\left(X,\hat{\theta}\right)\left(\tilde{f}-\Delta\right)$,
recall the inequalities (\ref{eq:Delta bound}) and $\left\Vert \Phi^{\prime}\left(X,\hat{\theta}\right)\right\Vert \leq\upsilon_{1}\left\Vert X\right\Vert +\upsilon_{2}$.
Using these inequalities, it can be shown that $\alpha_{3}\tilde{\theta}^{\top}\Phi^{\prime\top}\left(X,\hat{\theta}\right)\left(\tilde{f}-\Delta\right)\leq\rho_{4}\left(\left\Vert z\right\Vert \right)\left\Vert z\right\Vert \left\Vert \tilde{\theta}\right\Vert +\alpha_{3}\bar{\varepsilon}\left(\upsilon_{1}\overline{x_{d}}+\upsilon_{1}\overline{\dot{x}_{d}}+\upsilon_{2}\right)\left\Vert \tilde{\theta}\right\Vert $,
where $\rho_{4}:\mathbb{R}_{\geq0}\to\mathbb{R}_{\geq0}$ is a strictly
increasing function defined as $\rho_{4}\left(\left\Vert z\right\Vert \right)=\alpha_{3}(\upsilon_{1}(\alpha_{1}+2)\left\Vert z\right\Vert +\upsilon_{1}\overline{x_{d}}+\upsilon_{1}\overline{\dot{x}_{d}}+\upsilon_{2})(\left\Vert z\right\Vert +\rho_{1}\left(\left\Vert z\right\Vert \right)\left\Vert z\right\Vert ^{2})+\alpha_{3}\upsilon_{1}(\alpha_{1}+2)\bar{\varepsilon}\left\Vert z\right\Vert $.
By Young's inequality, it follows that $\alpha_{3}\tilde{\theta}^{\top}\Phi^{\prime\top}\left(X,\hat{\theta}\right)\left(\tilde{f}-\Delta\right)\leq\frac{k_{\hat{\theta}}}{2}\left\Vert \tilde{\theta}\right\Vert ^{2}+\frac{\rho_{4}^{2}\left(\left\Vert z\right\Vert \right)\left\Vert z\right\Vert ^{2}}{k_{\hat{\theta}}}+\frac{2\left(\alpha_{3}\bar{\varepsilon}\left(\upsilon_{1}\overline{x_{d}}+\upsilon_{1}\overline{\dot{x}_{d}}+\upsilon_{2}\right)\right)^{2}}{k_{\hat{\theta}}}$.
Hence, $\dot{V}$ can be further upper-bounded as
\begin{eqnarray}
\dot{V} & \leq & -\left(k_{\min}-\rho\left(\left\Vert z\right\Vert \right)\right)\left\Vert z\right\Vert ^{2}+c\nonumber \\
 &  & -\left(\alpha_{3}-\frac{1}{2}\right)\tilde{\theta}^{\top}\Phi^{\prime\top}\left(X,\hat{\theta}\right)\Phi^{\prime}\left(X,\hat{\theta}\right)\tilde{\theta},\label{eq:Vdot 2}
\end{eqnarray}
where $k_{\min}\triangleq\min\{\alpha_{1},\frac{k_{r}}{2},\alpha_{2},\frac{k_{f}}{2},\frac{k_{\hat{\theta}}}{4}+\frac{\beta_{1}}{2\lambda_{\Gamma,\max}}\}$,
$\rho\left(\left\Vert z\right\Vert \right)=\frac{\rho_{1}^{2}\left(\left\Vert z\right\Vert \right)\left\Vert z\right\Vert ^{2}}{k_{r}}+\frac{\rho_{3}\left(\left\Vert z\right\Vert \right)}{2k_{f}}+\frac{\rho_{4}^{2}\left(\left\Vert z\right\Vert \right)}{k_{\hat{\theta}}}$,
and $c=\frac{\bar{\varepsilon}^{2}}{k_{r}}+\frac{\rho_{2}^{2}\left(0\right)+\rho_{2}\left(0\right)\rho_{2}^{\prime}\left(0\right)}{2k_{f}}+\frac{2\left(\alpha_{3}\bar{\varepsilon}\left(\upsilon_{1}\overline{x_{d}}+\upsilon_{1}\overline{\dot{x}_{d}}+\upsilon_{2}\right)\right)^{2}}{k_{\hat{\theta}}}+k_{\hat{\theta}}\bar{\theta}^{2}$.
The term $\chi$ denoting the radius of $\mathcal{D}$ is now explicitly
defined as $\chi\triangleq\bar{\rho}^{-1}\left(k_{\min}-\lambda_{3}-\rho\left(0\right)\right)$,
where $\bar{\rho}:\mathbb{R}_{\geq0}\to\mathbb{R}_{\geq0}$ is a strictly
increasing invertible function defined as $\bar{\rho}\left(\left\Vert z\right\Vert \right)\triangleq\rho\left(\left\Vert z\right\Vert \right)-\rho\left(0\right)$,
and $\lambda_{3}\in\mathbb{R}_{>0}$ is the desired rate of convergence.
To facilitate the subsequent analysis, the following gain condition
is introduced 
\begin{equation}
\min\left\{ k_{\min}-\lambda_{3}-\rho\left(\sqrt{\frac{\lambda_{2}c}{\lambda_{1}\lambda_{3}}}\right),\alpha_{3}-\frac{1}{2}\right\} >0.\label{eq:gain condition}
\end{equation}
Additionally, the set $\mathcal{S}\triangleq\{\zeta\in\rr^{4n+p}:\left\Vert \zeta\right\Vert \leq\sqrt{\frac{\lambda_{1}}{\lambda_{2}}\chi^{2}-\frac{c}{\lambda_{3}}}\}$
is defined to initialize $z$ in the subsequent analysis, where it
is shown that if $z(t_{0})\in\mathcal{S}\subset\mathcal{D}$, then
$z(t)$ exhibits exponential converge to a neighborhood of the origin
and does not escape $\mathcal{D}$. The following theorem states the
main result of this paper.
\begin{thm}
\label{thm:main} Let Assumptions \ref{assm:activation bounds}-\ref{assm:Loss Function Strictly Convex}
and the gain condition in (\ref{eq:gain condition}) hold. Then, for
the system in (\ref{eq: dynamic system}), the DNN-based controller
in (\ref{eq:control input}) and the composite adaptation law in (\ref{eq:adapt law})
ensures that $\left\Vert z(t)\right\Vert \leq\sqrt{\frac{\lambda_{2}}{\lambda_{1}}\left\Vert z(t_{0})\right\Vert ^{2}\mathrm{e}^{-\frac{\lambda_{3}}{\lambda_{2}}(t-t_{0})}+\frac{\lambda_{2}c}{\lambda_{1}\lambda_{3}}\left(1-\mathrm{e}^{-\frac{\lambda_{3}}{\lambda_{2}}(t-t_{0})}\right)}$
for all $t\in[t_{0},\infty)$, provided that $\left\Vert z\left(t_{0}\right)\right\Vert \in\mathcal{S}$. 
\end{thm}
\begin{IEEEproof}
Due to the facts that $c>0$ and $\lambda_{1}\leq\lambda_{2}$ from
(\ref{eq:Lyap bounds}), it follows from the definitions of $\mathcal{S}$
and $\mathcal{D}$ that $\mathcal{S}\subset\mathcal{D}$. As a result,
$z(t_{0})\in\mathcal{S}$ implies $z(t_{0})\in\mathrm{int}\left(\mathcal{D}\right),$
where the notation $\mathrm{int}\left(\cdot\right)$ denotes the interior.
Therefore, because the solution $t\mapsto z(t)$ is continuous\footnote{Continuous solutions exists over some time-interval for systems satisfying
Caratheodory existence conditions. According to Caratheodory conditions
for the system $\dot{y}=f(y,t)$, $f$ should be locally bounded,
continuous in $y$ for each fixed $t$, and measurable in $t$ for
each fixed $y$ \cite[Ch.2, Theorem 1.1]{Coddington.Levinson1955}.
The dynamics in $\dot{z}$ satisfy the Caratheodory conditions.}, there exists a time-interval $\mathcal{I}\triangleq[t_{0},t_{1}]$
such that $z(t)\in\mathcal{D}$ for all $t\in\mathcal{I}$. In the
subsequent stability analysis, we analyze the convergence properties
of the solutions and also establish that $\mathcal{I}$ can be extended
to $[t_{0},\infty)$. Consider the candidate Lyapunov function in
(\ref{eq:Lyap}). Then, using (\ref{eq:Lyap bounds}) and (\ref{eq:Vdot 2}),
when the gain condition in (\ref{eq:gain condition}) is satisfied,
$\dot{V}$ can be upper-bounded as
\begin{eqnarray}
\dot{V} & \leq & -\frac{\lambda_{3}}{\lambda_{2}}V+c,\label{eq:Vdot 3-1}
\end{eqnarray}
for all $t\in\mathcal{I}$. Solving the differential inequality in
(\ref{eq:Vdot 3-1}) over the time-interval $\mathcal{I}$ yields
\begin{equation}
V\left(z(t)\right)\leq V\left(z(t_{0})\right)\mathrm{e}^{-\frac{\lambda_{3}}{\lambda_{2}}(t-t_{0})}+\frac{\lambda_{2}c}{\lambda_{3}}\left(1-\mathrm{e}^{-\frac{\lambda_{3}}{\lambda_{2}}(t-t_{0})}\right),\label{eq:V UUB Bound}
\end{equation}
for all $t\in\mathcal{I}$. Then applying (\ref{eq:Lyap bounds})
to (\ref{eq:V UUB Bound}) yields
\begin{equation}
\left\Vert z(t)\right\Vert \leq\sqrt{\frac{\lambda_{2}}{\lambda_{1}}\left\Vert z(t_{0})\right\Vert ^{2}\mathrm{e}^{-\frac{\lambda_{3}}{\lambda_{2}}(t-t_{0})}+\frac{\lambda_{2}c}{\lambda_{1}\lambda_{3}}\left(1-\mathrm{e}^{-\frac{\lambda_{3}}{\lambda_{2}}(t-t_{0})}\right)},\label{eq:UUB Bound}
\end{equation}
for all $t\in\mathcal{I}$. It remains to be shown that $\mathcal{I}$
can be extended to $[t_{0},\infty)$. Assume for the sake of contradiction
that $\mathcal{I}$ has to be bounded, i.e., the escape time $t_{1}$
is finite. Equivalently, there exists $t_{1}$ for which there does
not exist $t_{2}>t_{1}$ such that $z(t)\in\mathcal{D}$ for all $t\in[t_{1},t_{2}]$.
Substituting $t=t_{1}$ into (\ref{eq:UUB Bound}) yields $\left\Vert z(t_{1})\right\Vert \leq\sqrt{\frac{\lambda_{2}}{\lambda_{1}}\left\Vert z(t_{0})\right\Vert ^{2}\mathrm{e}^{-\frac{\lambda_{3}}{\lambda_{2}}(t_{1}-t_{0})}+\frac{\lambda_{2}c}{\lambda_{1}\lambda_{3}}\left(1-\mathrm{e}^{-\frac{\lambda_{3}}{\lambda_{2}}(t_{1}-t_{0})}\right)}<\sqrt{\frac{\lambda_{2}}{\lambda_{1}}\left\Vert z(t_{0})\right\Vert ^{2}+\frac{\lambda_{2}c}{\lambda_{1}\lambda_{3}}}$.
Because $z(t_{0})\in\mathcal{S}$, it follows that $\left\Vert z(t_{0})\right\Vert ^{2}\leq\frac{\lambda_{1}}{\lambda_{2}}\chi^{2}-\frac{c}{\lambda_{3}}$,
implying $\left\Vert z(t_{1})\right\Vert <\chi$, and hence $z(t_{1})\in\mathrm{int}\left(\mathcal{D}\right)$.
Therefore, by continuity of $z$, there exists $t_{2}>t_{1}$ such
that such that $z(t)\in\mathcal{D}$ for all $t\in[t_{1},t_{2}]$,
violating the assumption made for contradiction. Therefore, by contradiction,
$\mathcal{I}$ can be extended to the interval $[t_{0},\infty)$. 
\end{IEEEproof}
\begin{rem}
\label{rem:PE performance}Since $\lambda_{3}=\min\{\alpha_{1},\frac{k_{r}}{2},\alpha_{2},\frac{k_{f}}{2},\frac{k_{\hat{\theta}}}{4}+\frac{\beta_{1}}{2\lambda_{\Gamma,\max}}\}$,
the gains $\alpha_{1},\alpha_{2},k_{r},k_{f}$, and $k_{\hat{\theta}}$
can be selected to be sufficiently high such that $\lambda_{3}=\frac{k_{\hat{\theta}}}{4}+\frac{\beta_{1}}{2\lambda_{\Gamma,\max}}$.
Since $\beta_{1}$ is positive under the PE condition as mentioned
in Remark \ref{rem:PE property}, a larger value for $\lambda_{3}$
is obtained, which implies faster exponential convergence to a smaller
neighborhood of the origin. When the PE condition does not hold, convergence
is not guaranteed; however, the gain $k_{\hat{\theta}}$, which is
based on the sigma modification technique in \cite[Sec. 8.4.1]{Ioannou1996},
ensures boundedness of all states. However, selecting a high value
for $k_{\hat{\theta}}$ can deteriorate tracking and parameter estimation
performance since it yields a higher value for $c$. 
\end{rem}
\begin{rem}
\label{rem:gains}The gains have the following tuning interpretations:
$\alpha_{1}$ controls tracking error convergence rate in (\ref{eq:r});
$k_{r}$ provides robustness against disturbances in (\ref{eq:rdot 3});
$k_{f},\alpha_{2}$ govern observer convergence in (\ref{eq:observer_2})-(\ref{eq:r tilde dot});
and $k_{\hat{\theta}}$ ensures boundedness without PE per Remark
\ref{rem:PE performance}. In practice, gains are initialized conservatively
and increased while verifying (\ref{eq:gain condition}). High observer
gains are recommended to ensure the state-derivative estimation operates
at a faster timescale than the control and adaptation dynamics. 
\end{rem}

\section{\label{sec:Simulations}Simulations}

To demonstrate the performance of the developed method, comparative
simulations are performed on two different systems, i.e. a two-link
manipulator and a UUV.\footnote{Codes are available at: https://github.com/patilomkarsudhir/Composite-Adaptive-Lyapunov-Based-Deep-Neural-Network}

\subsection{\label{subsec:Two-Link-Manipulator}Two Link Manipulator}

To demonstrate the performance of the developed composite adaptive
Lb-DNN, comparative simulations are performed on the two-link robot
manipulator model (see Appendix IX.1.1 for the dynamics) in \cite{Queiroz.Hu.ea1997}
for 100 seconds. Baseline methods used for comparison include DNN-based
adaptive controller with tracking error-based adaptation law developed
in \cite{Patil.Le.ea2022}, an observer-based disturbance rejection
controller \cite{Han2009} (i.e., $u=g^{+}(x,\dot{x})\left(\ddot{x}_{d}-\left(\alpha_{1}+k_{r}\right)r+\left(\alpha_{1}^{2}-1\right)e-\hat{f}\right)$),
a nonlinear proportional-derivative (PD) controller $u=g^{+}(x,\dot{x})\left(\ddot{x}_{d}-\left(\alpha_{1}+k_{r}\right)r+\left(\alpha_{1}^{2}-1\right)e\right)$,
and nonlinear model predictive control (MPC). Complete implementation
details including MPC cost matrices, solver settings, and all gain
values are provided in Appendix IX.A.1 to ensure reproducibility.
The comparative simulation is performed using a fully-connected DNN
with 5 hidden layers and 5 neurons in each layer with hyperbolic tangent
activation functions. The DNN weights are initialized randomly from
the distribution $U(-0.5,0.5)$. For a realistic simulation, an additive
white Gaussian (AWG) measurement noise with a signal-to-noise ratio
of 50 dB is considered in all state measurements. 
\begin{table*}
\caption{\label{tab:Performance-Comparison 5x5}Robot Manipulator Performance
Comparison}

\centering{}%
\begin{tabular}{ccccc}
\hline 
{\scriptsize{}Adaptation Law} & {\scriptsize{}$\left\Vert e\right\Vert _{\textrm{RMS}}$ (deg)} & {\scriptsize{}$\left\Vert u\right\Vert _{\textrm{RMS}}$ (Nm)} & {\scriptsize{}$\begin{array}{c}
\textrm{Function}\\
\textrm{error}\\
\textrm{on-trajectory}\\
(\mathrm{rad/s^{2})}
\end{array}$} & {\scriptsize{}$\begin{array}{c}
\textrm{Mean }\\
\textrm{function}\\
\textrm{error on test}\\
\textrm{ data (\ensuremath{\mathrm{rad/s^{2})}}}
\end{array}$}\tabularnewline
\hline 
{\scriptsize{}Tracking Error-Based} & {\scriptsize{}0.629} & {\scriptsize{}10.100} & {\scriptsize{}0.430} & {\scriptsize{}1.215}\tabularnewline
\hline 
{\scriptsize{}Composite} & {\scriptsize{}0.308} & {\scriptsize{}7.962} & {\scriptsize{}0.131} & {\scriptsize{}0.260}\tabularnewline
\hline 
{\scriptsize Observer-Based} & {\scriptsize 0.310} & {\scriptsize 10.612} & {\scriptsize 0.204} & {\scriptsize N/A}\tabularnewline
\hline 
{\scriptsize Nonlinear PD} & {\scriptsize 3.142} & {\scriptsize 6.642} & {\scriptsize N/A} & {\scriptsize N/A}\tabularnewline
\hline 
{\scriptsize Nonlinear MPC} & {\scriptsize 1.101} & {\scriptsize 8.275} & {\scriptsize N/A} & {\scriptsize N/A}\tabularnewline
\hline 
\end{tabular}
\end{table*}

\begin{figure*}
\begin{centering}
\includegraphics[width=9cm]{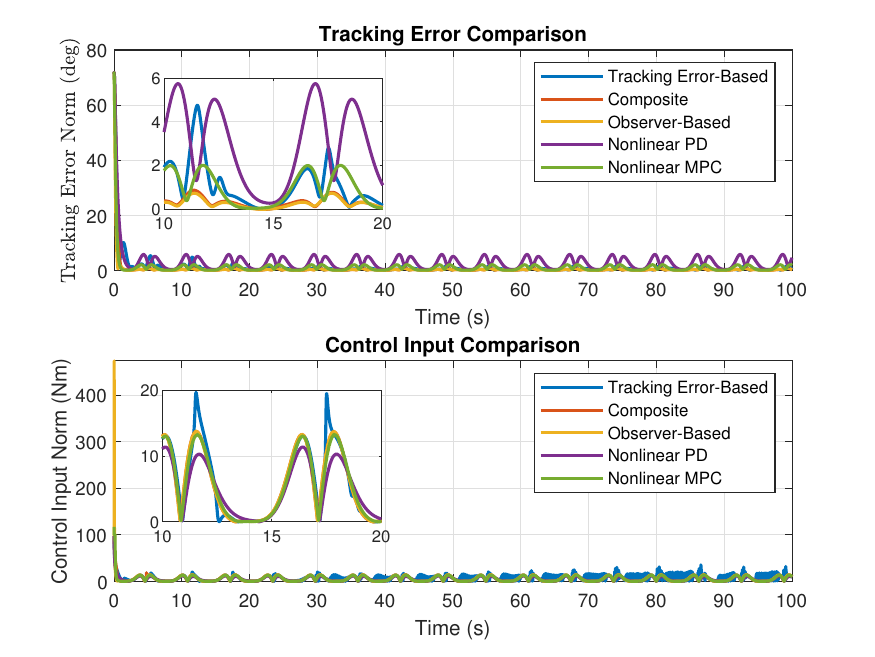}\includegraphics[width=9cm]{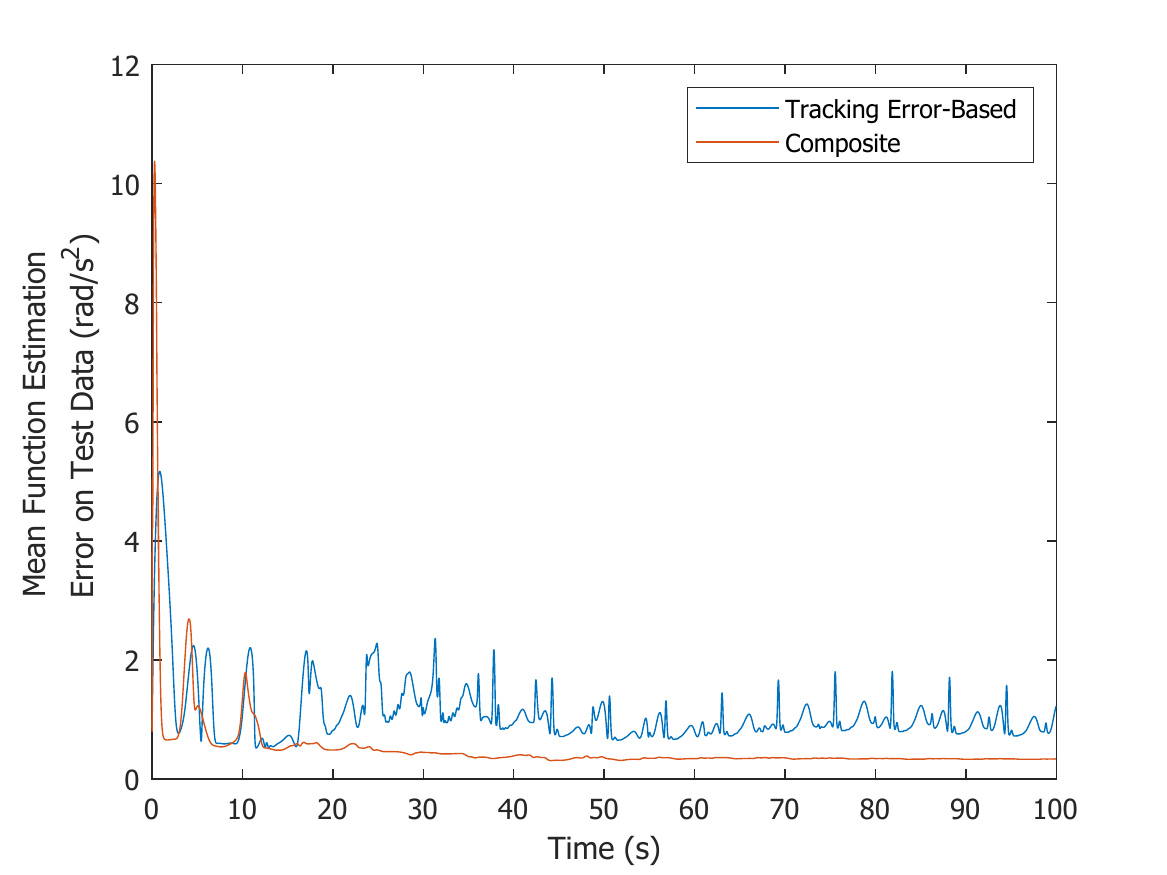}
\par\end{centering}
\caption{{\small{}\label{fig:Plots}Left: Comparative plots of the tracking
error norm and control input norm along the trajectory with the developed
and baseline controllers. A zoomed view during the time interval $[10,20]$
is added in each subplot for visual clarity. Right: Comparative plots
of the mean of function estimation error norm $\left\Vert f(x,\dot{x})-\Phi\left(X,\hat{\theta}\right)\right\Vert $
using tracking error-based adaptation and composite adaptation on
the test dataset.}}
\end{figure*}

To evaluate the tracking and drift compensation performance of the
developed and baseline methods, the root mean square (RMS) values
of the tracking error norm, denoted by $\left\Vert e\right\Vert _{\textrm{RMS}}$,
and function approximation error along the trajectory are calculated
in the steady state (i.e., in the interval {[}50,100{]} seconds).
The corresponding values are provided in Table \ref{tab:Performance-Comparison 5x5}.
Since the trajectory explored by the system essentially acts as a
training dataset for the DNNs, the RMS function approximation error
does not indicate whether the DNN model is overfit and how well the
model generalizes over unexplored data. Thus, to evaluate the performance
of the DNN beyond the trajectory, a test dataset involving 100 random
datapoints with values selected from the distribution $U(-0.25,0.25)$
is constructed, and the mean $\left\Vert f(x,\dot{x})-\Phi\left(X,\hat{\theta}\right)\right\Vert $
across all points in the dataset is evaluated. The value of the mean
$\left\Vert f(x,\dot{x})-\Phi\left(X,\hat{\theta}\right)\right\Vert $
on the test dataset at the end of each simulation (i.e., at $t=100$
seconds) is then used as a metric in Table \ref{tab:Performance-Comparison 5x5}
for comparing the generalization performance of each method. To evaluate
the control effort required by each controller throughout the transient
and steady states, the RMS values of the control input norm, $\left\Vert u\right\Vert $,
are calculated in the time-interval {[}0,100{]} seconds and provided
in Table \ref{tab:Performance-Comparison 5x5} for each method. As
evident from Table \ref{tab:Performance-Comparison 5x5} and Figure
\ref{fig:Plots}, the developed composite adaptive Lb-DNN significantly
improved the tracking performance compared to tracking error-based
adaptive Lb-DNN, nonlinear PD, and nonlinear MPC with comparable control
effort and approximately 50\%, 90\%, and 70\% improvements in $\left\Vert e\right\Vert _{\textrm{RMS}}$,
respectively. The tracking error-based adaptive Lb-DNN exhibited more
chattering in the control input due to measurement noise, which might
be because the update law involved a constant high adaptation gain.
In contrast, the composite update law has a decreasing gain due to
the least squares approach which mitigates noise amplification resulting
from the adaptation gain. Additionally, the tracking performance with
the observer-based disturbance rejection method is comparable to the
developed method, which is expected because the developed method also
used the observer-based estimate $\hat{f}$ to formulate the prediction
error. However, notice the increased control effort due to large overshoots
in the controller resulting from high gains in the state-derivative
observer in (\ref{eq:observer_2}). Although the developed composite
adaptation law also used the state-derivative estimates generated
by the high-gain observer, the state-derivative estimates are not
directly used in the control input. Using the state-derivative estimates
in the adaptation law did not cause as large overshoots in the control
input because the adaptation law involves an integrator that effectively
acts as a low pass filter on any overshoots in the state-derivative
estimates. Furthermore, note that the observer-based controller only
provides instantaneous estimates of $f$, due to the lack of a model.
Hence, it cannot be generalized for off-trajectory points, thus not
achieving the system identification objective. Additionally, despite
the fact that nonlinear MPC uses model knowledge, the developed method
achieved improved tracking with reduced control effort compared to
nonlinear MPC.

The evolution of the mean function approximation error on the aforementioned
test dataset is shown on the right in Figure \ref{fig:Plots}. The
mean function approximation error with the composite method on the
test dataset initially overshot followed by oscillatory behavior during
the initial 10 seconds. Such a behavior is expected since the combined
system goes through the initial transients, and the online data in
the first few seconds based on which the DNN has learned is limited.
However, after 10 s, the composite adaptation law exhibited a consistent
decrease in the mean function approximation error, unlike the tracking
error-based adaptation law. As a result, the composite adaptation
law achieved 72.04\% improvement in the final value of mean function
approximation error.

\subsubsection{Ablation Study}

\begin{table*}
\caption{\label{tab:Performance-Comparison-1}Performance Comparison}

\centering{}%
\begin{tabular}{ccccccc}
\hline 
\multicolumn{2}{c}{{\scriptsize{}Architecture}} & {\scriptsize{}Adaptation Law} & {\scriptsize{}$\left\Vert e\right\Vert _{\textrm{RMS}}$ (deg)} & {\scriptsize{}$\left\Vert u\right\Vert _{\textrm{RMS}}$ (Nm)} & {\scriptsize{}$\begin{array}{c}
\textrm{Function}\\
\textrm{error}\\
\textrm{on-trajectory}\\
(\mathrm{rad/s^{2})}
\end{array}$} & {\scriptsize{}$\begin{array}{c}
\textrm{Final mean }\\
\textrm{function}\\
\textrm{error on test}\\
\textrm{ data (\ensuremath{\mathrm{rad/s^{2})}}}
\end{array}$}\tabularnewline
\hline 
{\scriptsize{}Layers} & {\scriptsize{}Neurons} &  &  &  &  & \tabularnewline
\hline 
\multirow{3}{*}{{\scriptsize{}3}} & \multirow{3}{*}{{\scriptsize{}3}} & {\scriptsize{}Tracking Error-Based} & {\scriptsize{}0.731} & {\scriptsize{}7.484} & {\scriptsize{}0.430} & {\scriptsize{}0.954}\tabularnewline
\cline{3-7}
 &  & {\scriptsize{}Composite} & {\scriptsize{}0.315} & {\scriptsize{}7.944} & {\scriptsize{}0.138} & {\scriptsize{}0.140}\tabularnewline
\cline{3-7}
 &  & {\scriptsize{}\% Decrease} & {\scriptsize{}56.97} & {\scriptsize{}-6.14} & {\scriptsize{}67.77} & {\scriptsize{}85.29}\tabularnewline
\hline 
\multirow{3}{*}{{\scriptsize{}4}} & \multirow{3}{*}{{\scriptsize{}3}} & {\scriptsize{}Tracking Error-Based} & {\scriptsize{}3.338} & {\scriptsize{}7.021} & {\scriptsize{}1.636} & {\scriptsize{}0.725}\tabularnewline
\cline{3-7}
 &  & {\scriptsize{}Composite} & {\scriptsize{}0.338} & {\scriptsize{}7.957} & {\scriptsize{}0.154} & {\scriptsize{}0.325}\tabularnewline
\cline{3-7}
 &  & {\scriptsize{}\% Decrease} & {\scriptsize{}89.87} & {\scriptsize{}-13.33} & {\scriptsize{}90.58} & {\scriptsize{}55.024}\tabularnewline
\hline 
\multirow{3}{*}{{\scriptsize{}4}} & \multirow{3}{*}{{\scriptsize{}4}} & {\scriptsize{}Tracking Error-Based} & {\scriptsize{}0.685} & {\scriptsize{}7.731} & {\scriptsize{}0.426} & {\scriptsize{}0.759}\tabularnewline
\cline{3-7}
 &  & {\scriptsize{}Composite} & {\scriptsize{}0.309} & {\scriptsize{}7.957} & {\scriptsize{}0.132} & {\scriptsize{}0.154}\tabularnewline
\cline{3-7}
 &  & {\scriptsize{}\% Decrease} & {\scriptsize{}54.85} & {\scriptsize{}-2.87} & {\scriptsize{}68.93} & {\scriptsize{}79.64}\tabularnewline
\hline 
\multirow{3}{*}{{\scriptsize{}5}} & \multirow{3}{*}{{\scriptsize{}5}} & {\scriptsize{}Tracking Error-Based} & {\scriptsize{}0.664} & {\scriptsize{}7.800} & {\scriptsize{}0.395} & {\scriptsize{}1.22}\tabularnewline
\cline{3-7}
 &  & {\scriptsize{}Composite} & {\scriptsize{}0.307} & {\scriptsize{}7.955} & {\scriptsize{}0.131} & {\scriptsize{}0.342}\tabularnewline
\cline{3-7}
 &  & {\scriptsize{}\% Decrease} & {\scriptsize{}53.69} & {\scriptsize{}-1.99} & {\scriptsize{}66.87} & {\scriptsize{}72.04}\tabularnewline
\hline 
\multirow{3}{*}{{\scriptsize{}5}} & \multirow{3}{*}{{\scriptsize{}10}} & {\scriptsize{}Tracking Error-Based} & {\scriptsize{}0.351} & {\scriptsize{}7.940} & {\scriptsize{}0.192} & {\scriptsize{}1.010}\tabularnewline
\cline{3-7}
 &  & {\scriptsize{}Composite} & {\scriptsize{}0.308} & {\scriptsize{}7.959} & {\scriptsize{}0.130} & {\scriptsize{}0.110}\tabularnewline
\cline{3-7}
 &  & {\scriptsize{}\% Decrease} & {\scriptsize{}12.41} & {\scriptsize{}-0.235} & {\scriptsize{}32.13} & {\scriptsize{}89.07}\tabularnewline
\hline 
\multirow{3}{*}{{\scriptsize{}10}} & \multirow{3}{*}{{\scriptsize{}10}} & {\scriptsize{}Tracking Error-Based} & {\scriptsize{}0.584} & {\scriptsize{}8.826} & {\scriptsize{}1.330} & {\scriptsize{}2.624}\tabularnewline
\cline{3-7}
 &  & {\scriptsize{}Composite} & {\scriptsize{}0.307} & {\scriptsize{}7.965} & {\scriptsize{}0.130} & {\scriptsize{}0.206}\tabularnewline
\cline{3-7}
 &  & {\scriptsize{}\% Decrease} & {\scriptsize{}47.34} & {\scriptsize{}9.75} & {\scriptsize{}90.22} & {\scriptsize{}92.15}\tabularnewline
\cline{3-7}
\end{tabular}
\end{table*}

An ablation study is performed to demonstrate the performance of the
developed method for various DNN architectures mentioned in Table
\ref{tab:Performance-Comparison-1}. The same set of gains are used
and the weights are randomly initialized from the distribution $U(-0.5,0.5)$.
As evident from the percentage decrease in Table \ref{tab:Performance-Comparison-1},
the developed composite adaptation law significantly improves the
tracking, drift compensation, and generalization performance of the
DNN across all DNN architecture with a comparable control effort.
Notably, although all DNN architectures yielded acceptable performance
(i.e., with $\left\Vert e\right\Vert _{\textrm{RMS}}$ less than 0.5
deg) with the developed composite adaptive Lb-DNN controller, no conclusive
trend was obtained to comment on the selection of appropriate size
for the DNN for this application. Importantly, using DNN of a greater
size did not affect the control effort.

\subsection{\label{subsec:Unmanned-Underwater-Vehicle}Unmanned Underwater Vehicle}

\begin{table*}
\caption{\label{tab:UUV Table}UUV Performance Comparison}

\centering{}{\scriptsize{}}%
\begin{tabular}{cccccc}
\hline 
 & {\tiny{}$e$-based} & {\tiny{}Composite} & {\tiny{}Observer-based} & {\tiny NMPC} & {\tiny NPD}\tabularnewline
\hline 
\hline 
{\tiny{}RMS position tracking error norm (m)} & {\tiny{}0.201} & {\tiny{}0.152} & {\tiny{}0.158} & {\tiny 0.254} & {\tiny 0.186}\tabularnewline
\hline 
{\tiny{}RMS angular tracking error norm (rad)} & {\tiny{}0.037} & {\tiny{}0.012} & {\tiny{}0.024} & {\tiny 0.054} & {\tiny 0.028}\tabularnewline
\hline 
{\tiny{}RMS linear control input norm (N)} & {\tiny{}0.069} & {\tiny{}0.065} & {\tiny{}0.126} & {\tiny 0.128} & {\tiny 0.067}\tabularnewline
\hline 
{\tiny{}RMS angular control input norm (Nm)} & {\tiny{}0.041} & {\tiny{}0.035} & {\tiny{}0.072} & {\tiny 0.036} & {\tiny 0.032}\tabularnewline
\hline 
{\tiny{}RMS linear dynamics estimation error norm ($\mathrm{m}/\mathrm{s}^{2}$)} & {\tiny{}4.370} & {\tiny{}2.585} & {\tiny{}19.423} & {\tiny N/A} & {\tiny N/A}\tabularnewline
\hline 
{\tiny{}RMS angular dynamics estimation error norm ($\mathrm{rad}/\mathrm{s}^{2}$)} & {\tiny{}2.058} & {\tiny{}1.408} & {\tiny{}3.021} & {\tiny N/A} & {\tiny N/A}\tabularnewline
\hline 
\end{tabular}
\end{table*}

\begin{figure}
\begin{centering}
\includegraphics[width=9cm]{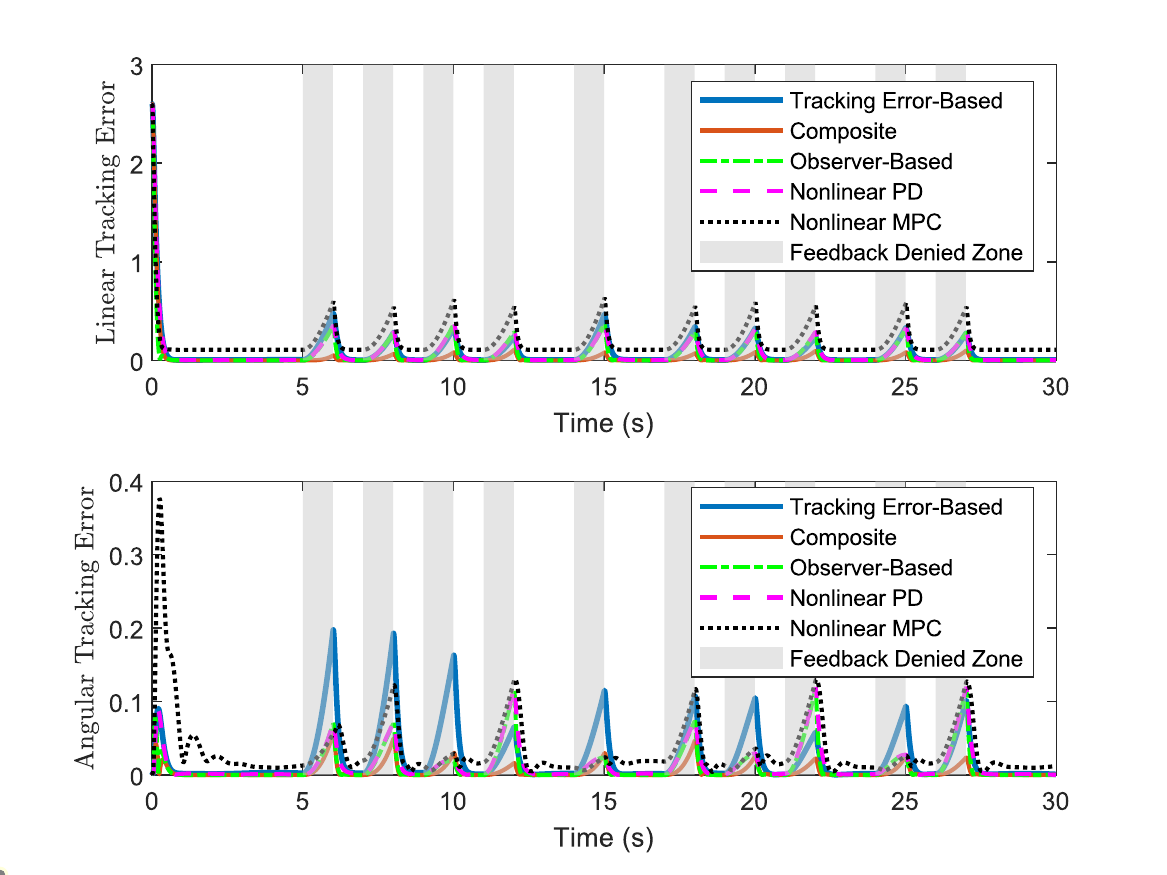}
\par\end{centering}
\caption{{\small{}\label{fig:UUV 3D Trajectory}Comparative plots of the linear
tracking error norm (m) and angular tracking error norm (rad) for
the UUV. The time intervals corresponding to the feedback denied zones
are marked in grey patches.}}
\end{figure}

\begin{figure*}
\begin{centering}
\includegraphics[width=9cm]{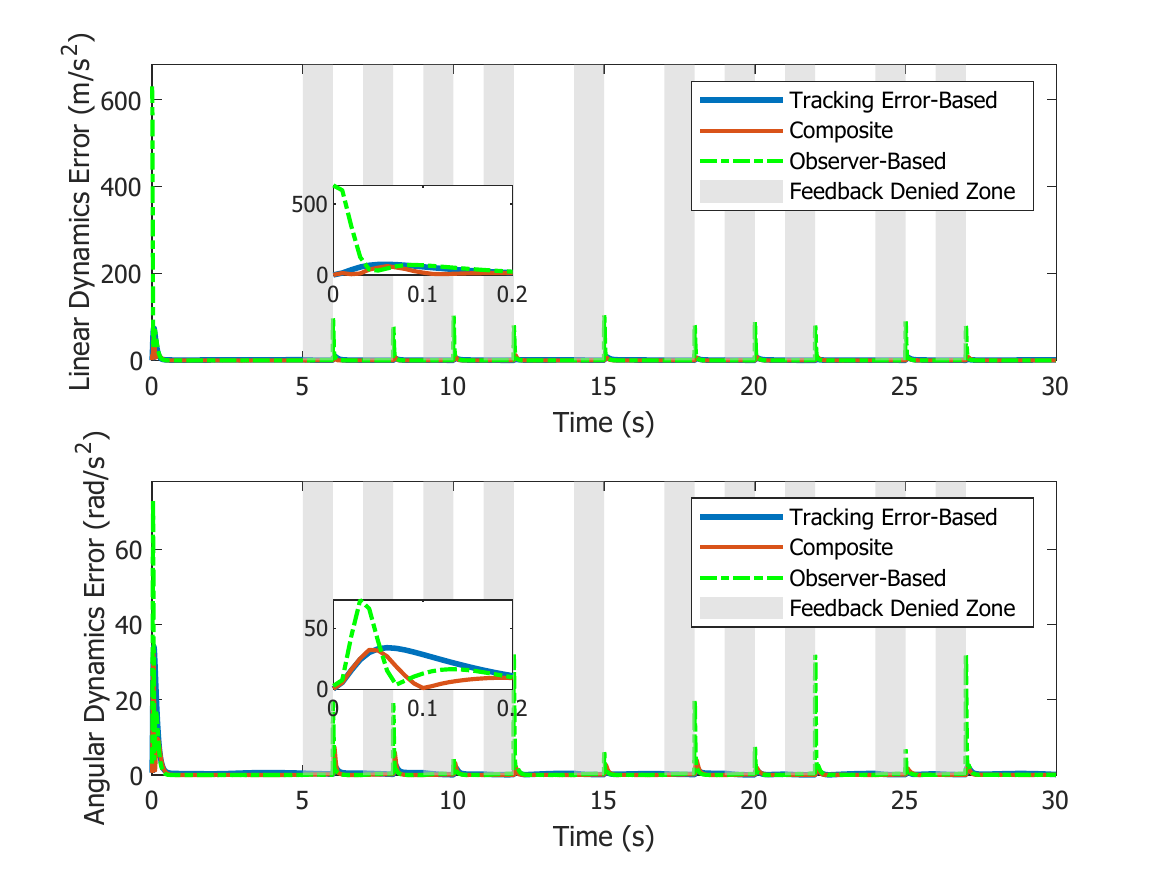}\includegraphics[width=9cm]{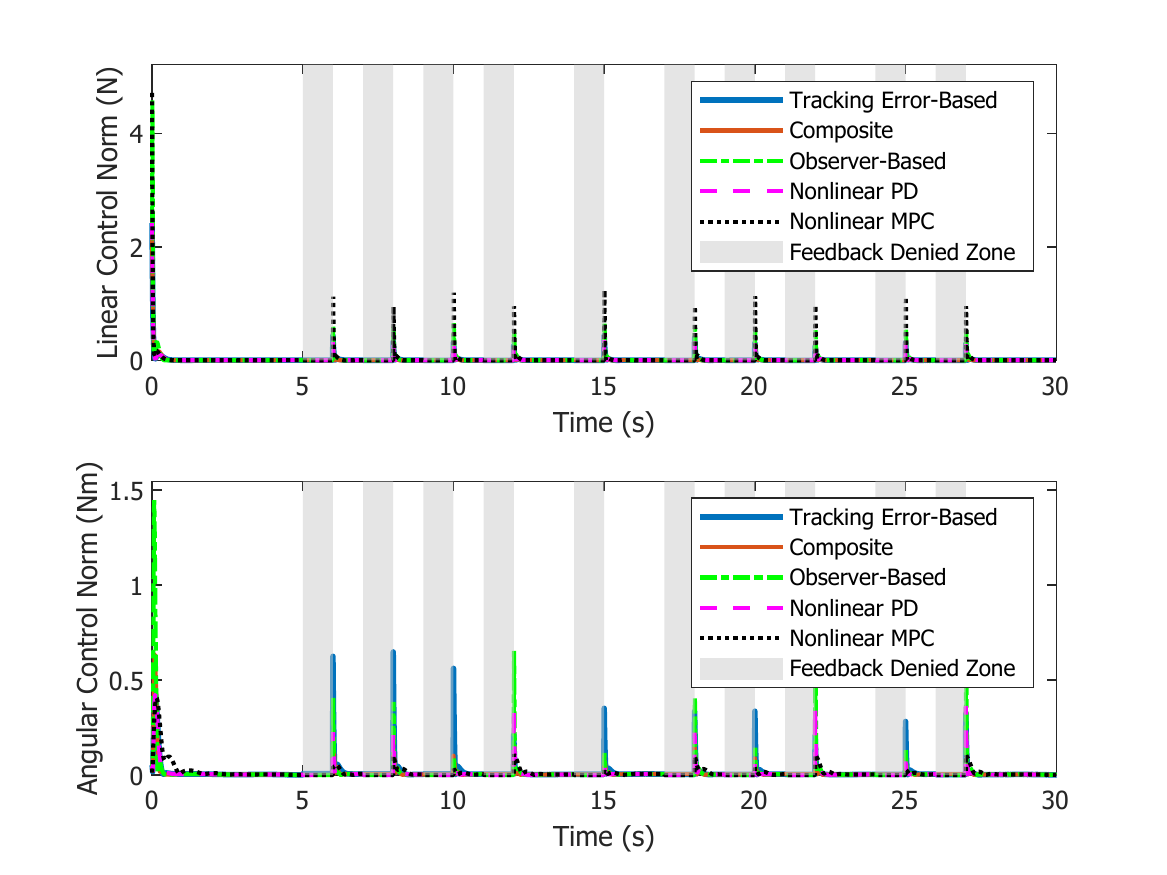}
\par\end{centering}
\caption{\label{fig:function error}{\small Left: Comparative plots of the estimation
error norm (i.e., $\left\Vert f(x,\dot{x})-\Phi\left(X,\hat{\theta}\right)\right\Vert $
during feedback availability and $\left\Vert f(x,\dot{x})-\Phi\left(X_{d}{(t)},\hat{\theta}\right)\right\Vert $
during loss of feedback) for the UUV with the developed and baseline
methods. Right: Comparative plots of the linear and angular control
input norms for the UUV. Zoomed views during the time interval $[0,0.2]$
seconds are added for visual clarity. }}
\end{figure*}

Comparative simulation results are also provided for the UUV system
(see Appendix IX.2 for dynamics) from \cite{Fischer.Hughes.ea2014}
using the composite adaptive Lb-DNN under intermittent loss of feedback,
with the same baselines as in Subsection \ref{subsec:Two-Link-Manipulator}.
Since the DNN identifies the system dynamics, the identified DNN could
be used to predict the uncertainty when the state feedback is intermittently
lost. Let $i\in\mathbb{Z}_{\geq0}$ denote the time index such that
the state feedback is available in the time interval $[t_{2i},t_{2i+1})$
and unavailable in the time interval $[t_{2i+1},t_{2i+2})$ for all
$i\in\mathbb{Z}_{\geq0}$. During the time interval $[t_{2i},t_{2i+1})$,
when the feedback is available, the control and adaptation laws in
(\ref{eq:control input}) and (\ref{eq:adapt law}) are used for all
$i\in\mathbb{Z}_{\geq0}$. However, during the time interval $[t_{2i+1},t_{2i+2})$
when the state feedback is unavailable, an open-loop controller is
developed as $u=g^{+}\left(x_{d}{(t)},\dot{x}_{d}{(t)}\right)\left(\ddot{x}_{d}{(t)}-\Phi\left(X_{d}{(t)},\hat{\theta}(t_{2i+1})\right)\right).$
The reader is referred to Appendix X for sufficient dwell-time conditions
and stability analysis under which the system can stably operate under
intermittent loss of feedback. For both the composite and tracking
error-based adaptive Lb-DNN methods, a fully-connected DNN with 5
hidden layers with 5 neurons in each layer with hyperbolic tangent
activation function was used. During feedback unavailability, the
observer-based disturbance rejection and nonlinear PD controllers
were designed to be $u=g^{+}\left(x_{d}{(t)},\dot{x}_{d}{(t)}\right)\ddot{x}_{d}{(t)}$,
with $\dot{\hat{x}},\dot{\hat{f}}=0$ for the observer-based method,
and the nonlinear MPC was designed using model predictions propagated
forward over the horizon treating $x_{d}$ as the current state. To
simulate the performance of the system under intermittent loss of
feedback, each simulation was performed for 30 seconds where the feedback
was made unavailable for the time intervals $[5,6)$, $[7,8)$, $[9,10)$,
$[11,12)$, $[14,15)$, $[17,18)$, $[19,20)$, $[21,22)$, $[24,25)$,
and $[26,27)$ seconds, respectively. For a realistic simulation,
an AWG measurement noise with a signal-to-noise ratio of 50 dB is
considered in all state measurements.

Figure \ref{fig:UUV 3D Trajectory} shows comparative plots of linear
tracking error and angular tracking error norms. The developed method
outperforms the baseline methods in tracking the reference trajectory,
especially during the loss of feedback as also evident from Table
\ref{tab:UUV Table}. Additionally, Figure \ref{fig:function error}
shows the comparative plot of the linear and angular dynamics (function)
estimation error norms on the left and control input norms on the
right. Since tracking error-based adaptation does not involve guarantees
on parameter estimation, the resulting predictions quickly diverge
during absence of feedback due to model identification errors. However,
the composite adaptive Lb-DNN controller can better predict and compensate
for the drift dynamics under the absence of state feedback. Additionally,
when feedback is available, the state-derivative observer-based approach
can yield a tracking performance comparable to the DNN-based controllers
since it essentially involves a robust high-gain approach. However,
the tracking performance degrades significantly in the absence of
feedback using the observer-based approach when compared to the composite
adaptive Lb-DNN controller. Furthermore, the observer-based approach
requires significantly higher control effort as compared to both of
the DNN-based adaptive controllers due to reasons discussed in Subsection
\ref{subsec:Two-Link-Manipulator}. Additionally, despite the fact
that nonlinear MPC uses model knowledge, the developed method achieved
improved tracking with reduced control effort compared to nonlinear
MPC.

\section{\label{sec:Conclusion}Conclusion}

A composite adaptive Lb-DNN is developed for simultaneous online system
identification and control, using the Jacobian of the DNN, the tracking
error, and a prediction error based on a novel formulation using a
dynamic state-derivative observer. A Lyapunov-based stability analysis
guarantees the tracking, observer, and parameter estimation errors
are UUB, with tighter bounds on these errors when the DNN's Jacobian
satisfies the PE condition. Comparative simulation results demonstrate
a significant improvement in tracking, function estimation and generalization
capabilities with the developed method in comparison to the tracking
error-based Lb-DNN in \cite{Patil.Le.ea2022} and observer-based disturbance
rejection controller as baseline methods.

\section{Limitations and Scope for Future Work}

The persistence of excitation condition for identifying the parameters
is restrictive as there are challenges in verifying it online. For
linear regression, recent developments in the adaptive control literature
such as \cite{Chowdhary.Yucelen.ea2012,Roy.Bhasin.ea2017,Pan.Yu2018,Ortega.Aranovskiy.2021}
provide parameter estimation methods that guarantee parameter convergence
under excitation conditions weaker than PE. All of these methods involve
some form of regression extension by storing the history of the regression
in memory over an interval of time. However, these methods are restricted
to linear regression and have not been explored for NIP models such
as DNNs yet. Thus, insights from this paper may be used in future
work to develop adaptation laws for DNNs that yield parameter estimation
guarantees under excitation conditions weaker than PE. Furthermore,
extensions of the developed online system identification approach
in optimization-based control paradigms such as MPC and reinforcement
learning can be explored. Moreover, future research efforts can also
investigate how to combine the developed method with control barrier
functions to satisfy state and input constraints.

\bibliographystyle{ieeetr}
\bibliography{encr,master,ncr,new_entries}

\section*{Appendices}

\section{Related Work}

\subsection{On Neural Network-Based Adaptive Control}

Classical results \cite{Lewis1996,Lewis1996a,Ge2002} develop adaptive
controllers for neural networks with a single hidden layer, where
online updates are performed for the input and output layer weights.
In recent results \cite{Joshi.Virdi.ea.2020b,Sun.Greene.ea2021,Le.Greene.ea2021},
adaptive controllers were developed for DNNs. In these results, the
outer-layer weights of the DNN are updated in real-time using Lyapunov-based
adaptation laws, whereas the inner-layer weights are updated either
using iterative batch updates on discrete intervals of time \cite{Joshi.Virdi.ea.2020b,Sun.Greene.ea2021,Le.Greene.ea2021},
or using a modular design \cite{Le.Greene.ea2021}. Since the inner-layer
weight updates in \cite{Joshi.Virdi.ea.2020b,Sun.Greene.ea2021,Le.Greene.ea2021}
happen using batch updates, the updates are essentially performed
offline. In \cite{Le.Greene.ea2021}, the weights are updated online
but the update laws are selected arbitrarily and not by a stability-driven
approach. In \cite{Patil.Le.ea2022}, Lyapunov-based adaptation laws
are developed for all layers of a fully-connected DNN (i.e., so-called
Lb-DNN methods). Since the control and adaptation laws are derived
from a Lyapunov-based stability analysis, the development is guaranteed
to ensure stability of the closed-loop system. More recent Lb-DNN
results develop Lyapunov-based adaptation laws for more complex architectures,
specifically, deep residual networks (ResNets) \cite{Patil.Le.ea.2022}
and long short-term memory (LSTM) networks \cite{Griffis.Patil.ea23_2}.
However, as stated in the manuscript, the updates are based solely
on tracking error feedback and are primarily meant to achieve tracking
error convergence. These results do not achieve guarantees on parameter
estimation and system identification.

\subsection{On Composite Adaptive Control}

The classical result in \cite{Slotine1989} develops adaptive controllers
with a composite adaptation law that includes both tracking and prediction
errors for nonlinear systems with linear-in-parameters (LIP) uncertainties.
The result in \cite{Slotine1989} constructs a form of the prediction
error using the swapping technique (also known as input or torque
filtering), where a low-pass filter is applied on both sides of the
dynamics to eliminate the unknown state-derivative term. For a brief
illustration of the swapping technique in \cite{Slotine1989}, consider
the system 
\[
\dot{x}=Y(x)\theta+u,
\]
where $Y(x)$ is the regressor, $\theta$ is the vector of unknown
parameters, and $u$ is the control input. If the state-derivative
$\dot{x}$ could be measured, the system can be expressed in terms
of the linear regression equation $\dot{x}-u=Y(x)\theta$. Then the
corresponding prediction error with an adaptive estimate $\hat{\theta}$
could be developed as $\epsilon=\dot{x}-u-Y(x)\hat{\theta}$, which
can be expressed linearly in terms of the parameter estimation error
$\tilde{\theta}=\theta-\hat{\theta}$ as $\epsilon=Y(x)\tilde{\theta}$.
However, $\dot{x}$ measurements are typically either unavailable
or extremely noisy. To avoid using state-derivative information, \cite{Slotine1989}
applied a low-pass filter on both sides of the dynamics, which results
in the filtered regression 
\begin{eqnarray*}
e^{-\beta t}*\left(\dot{x}-u\right) & = & \left(e^{-\beta t}*Y(x)\right)\theta,
\end{eqnarray*}
where $*$ denotes the convolutional integral operation (i.e., $a(t)*b(t)=\int_{0}^{t}a(t-\tau)b(\tau)d\tau$)
and $e^{-\beta t}$ is the impulse response of the low-pass filter
with a positive constant decay rate $\beta$. Since $e^{-\beta t}*\dot{x}=x(t)-x(0)e^{-\beta t}+\beta e^{-\beta t}*x$,
the filtered regression can be expressed as 
\[
x(t)-x(0)e^{-\beta t}+\beta e^{-\beta t}*x-e^{-\beta t}*u=\left(e^{-\beta t}*Y(x)\right)\theta,
\]
which is implementable without using state-derivative information.
The prediction error for the filtered regression can be developed
as 
\begin{eqnarray*}
\epsilon & = & x(t)-x(0)e^{-\beta t}+\beta e^{-\beta t}*x\\
 &  & -e^{-\beta t}*u-\left(e^{-\beta t}*Y(x)\right)\hat{\theta}\\
 & = & \left(e^{-\beta t}*Y(x)\right)\theta-\left(e^{-\beta t}*Y(x)\right)\hat{\theta}\\
 & = & \left(e^{-\beta t}*Y(x)\right)\tilde{\theta}.\\
 & = & Y_{f}\tilde{\theta},
\end{eqnarray*}
where $Y_{f}=\left(e^{-\beta t}*Y(x)\right)$. Since $\epsilon$ is
linear in $\tilde{\theta}$, a composite adaptation law can be developed
with a $Y_{f}^{\top}\epsilon$ term which would yield negative $\tilde{\theta}$
terms in the corresponding Lyapunov-based stability analysis. However,
yielding this form of $\epsilon$ using a filtered regression was
possible because the uncertainty $Y(x)\theta$ is linear in terms
of $\theta$, which allowed $\theta$ to be separable from $e^{-\beta t}*Y(x)$
in the filtered regression. If $Y(x)\theta$ is replaced by terms
nonlinear in $\theta$, such as the DNN-based approximation $\Phi(x,\theta)+\varepsilon(x)$,
applying a low pass filter on both sides would yield 
\[
e^{-\beta t}*\left(\dot{x}-u\right)=e^{-\beta t}*\left(\Phi(x,\theta)+\varepsilon(x)\right).
\]
Notice that $\theta$ is not separable from the convolutional integral
in the term $e^{-\beta t}*\Phi(x,\theta)$ since $\Phi$ is nonlinear
in $\theta$. As a result, the swapping technique from \cite{Slotine1989}
does not apply for nonlinear in parameter uncertainties such as DNNs.

Results in \cite{Patre2010b} introduce a robust integral of the sign
of the error (RISE)-based swapping technique to formulate the prediction
error and design composite adaptive controllers for LIP uncertainties
with additive disturbances. The RISE-based swapping technique is extended
in \cite{Patre2010a} for NN-based models, but the development is
restricted to single-hidden-layer NNs. Extending this for DNNs is
mathematically challenging due to their nested NIP structure. Moreover,
using RISE-based swapping requires additional RISE-based terms in
the control input, which can debilitate the learning performance of
the adaptive feedforward term. Notably, the results in \cite{Patre2010b}
and \cite{Patre2010a} only ensure asymptotic tracking error convergence,
and no guarantees are provided on the parameter estimates under the
persistence of excitation (PE) condition.

The recent result in \cite{OConnell.Shi.ea2022} developed a new learning
representation uncertainties involving a composited disturbance given
by $f(x,\dot{x},w)=\phi\left(x,\dot{x}\right)a\left(w\right)$, where
$\phi\left(\cdot\right)$ denotes a basis function that is learned
using a DNN and $a\left(w\right)$ denotes a set of linear parameters
accounting for an unknown disturbance time-varying disturbance $w$.
Since $f$ is linear in terms of $a$, the composite adaptive approach
from \cite{Slotine1989} is used to design an adaptation law $\dot{\hat{a}}$
to update the estimates of $a$ given by $\hat{a}$. To obtain a disturbance-invariant
representation of $\phi$ using DNNs, a domain adversarially invariant
meta-learning (DAIML) algorithm is developed to train the DNN offline.
To the best of our knowledge, this is the only existing work using
a composite adaptive approach in the context of deep learning-based
control. However, since the DNN learning $\phi\left(x,\dot{x}\right)$
has an NIP structure, the aforementioned challenges apply for constructing
a Lyapunov-based online adaptation law. 

\section{More Simulation Result Details}

All simulations were performed in MATLAB on a desktop with 64 GB RAM
and 13th Gen Intel Core i9-13900 @2.00 GHz processor.

\subsection{Two Link Manipulator}

\subsubsection{Dynamic Model}

The two-link robot manipulator was modeled by the uncertain Euler-Lagrange
dynamics 
\begin{align}
M\left(x\right)\ddot{x}+C\left(x,\dot{x}\right)\dot{x}+F\dot{x} & =u,\label{sim: two-link dynamics}
\end{align}
where $x\triangleq\left[x_{1},x_{2}\right]^{\top}\in\mathbb{R}^{2}$,
$\dot{x}\in\mathbb{R}^{2}$, and $\ddot{x}\in\mathbb{R}^{2}$ denote
the vector of angular position, velocity, and acceleration of joints,
respectively, $M\left(x\right)\in\mathbb{R}^{2\times2}$ represents
the inertia matrix, $C\left(x,\dot{x}\right)\in\mathbb{R}^{2\times2}$
represents the centripetal-Coriolis matrix, $F\in\mathbb{R}^{2\times2}$
represents friction effects, and $u\in\mathbb{R}^{2}$ denotes the
torque inputs. In (\ref{sim: two-link dynamics}), the dynamics were
modeled as \cite{Queiroz.Hu.ea1997} 
\begin{equation}
M\left(x\right)=\left[\begin{array}{cc}
p_{1}+2p_{3}c_{2}, & p_{2}+p_{3}c_{2}\\
p_{2}+p_{3}c_{2}, & p_{2}
\end{array}\right],\label{sim: M}
\end{equation}
\begin{equation}
C\left(x,\dot{x}\right)=\left[\begin{array}{cc}
-p_{3}s_{2}\dot{x}_{2}, & -p_{3}s_{2}\left(\dot{x}_{1}+\dot{x}_{2}\right)\\
p_{3}s_{2}\dot{x}_{1}, & 0
\end{array}\right],\label{sim: Vm}
\end{equation}
\begin{equation}
F=\left[\begin{array}{cc}
f_{1}, & 0\\
0, & f_{2}
\end{array}\right],\label{sim: F}
\end{equation}
where the short-hand notations $c_{2}$ and $s_{2}$ are defined as
$c_{2}\triangleq\cos\left(x_{2}\right)$ and $s_{2}\triangleq\sin\left(x_{2}\right)$
, respectively. The nominal parameters of the two-link robot model
in (\ref{sim: M})--(\ref{sim: F}) were $p_{1}=3.473\,\mathrm{kg\cdot m^{2}}$
, $p_{2}=0.196\,\mathrm{kg\cdot m^{2}}$, $p_{3}=0.242\,\mathrm{kg\cdot m^{2}}$,
$f_{1}=5.3\,\mathrm{Nm\cdot sec}$, and $f_{2}=1.1\,\mathrm{Nm\cdot sec}$.
The two-link manipulator dynamics can be expressed using Eq. (1) from
the manuscript with $f(x,\dot{x})=-M^{-1}\left(x\right)\left(C\left(x,\dot{x}\right)\dot{x}+F\dot{x}\right)$
and $g(x,\dot{x})=M^{-1}\left(x\right)$. The gains are selected as
$\alpha_{1}=5$, $\alpha_{2}=10$, $\alpha_{3}=20$, $\Gamma(0)=I,$
$k_{r}=5$, $k_{f}=10$, $k_{\hat{\theta}}=0.0001$, $\beta_{0}=10$,
and $\varkappa_{0}=2$. The states are initialized as $x(0)=[1,-1]^{\top}$
rad and $\dot{x}(0)=[0,0]^{\top}$ rad/s, the initial parameter estimate
$\hat{\theta}(0)$ is selected from the uniform distribution $U(-0.5,0.5)$,
and the desired trajectory is $x_{d}=0.25\exp(-\sin(t))[\sin(t),\cos(t)]^{\top}$
rad. The weights are randomly initialized from the distribution $U(-0.5,0.5)$.
Baseline methods used for comparison include DNN-based adaptive controller
with tracking error-based adaptation law developed in \cite{Patil.Le.ea2022},
an observer-based disturbance rejection controller \cite{Han2009}
(i.e., $u=g^{+}(x,\dot{x})\left(\ddot{x}_{d}-\left(\alpha_{1}+k_{r}\right)r+\left(\alpha_{1}^{2}-1\right)e-\hat{f}\right)$),
a nonlinear proportional-derivative (PD) controller $u=g^{+}(x,\dot{x})\left(\ddot{x}_{d}-\left(\alpha_{1}+k_{r}\right)r+\left(\alpha_{1}^{2}-1\right)e\right)$,
and nonlinear model predictive control (MPC). The baseline DNN-based
adaptive controller uses the tracking error-based adaptation law given
by \cite{Patil.Le.ea2022} 
\[
\dot{\hat{\theta}}=\mathrm{proj}\left(-k_{\hat{\theta}}\Gamma{(t_{0})}\hat{\theta}+\Gamma{(t_{0})}\Phi^{\prime\top}\left(X,\hat{\theta}\right)r\right)
\]
with a constant $\Gamma$ (unlike the developed method which uses
a time-varying $\Gamma$), where it was selected as $\Gamma=I$. For
a fair comparison, the set of gains common to the developed and baseline
methods were selected to be exactly the same. The nonlinear MPC was
designed to minimize the cost 
\begin{eqnarray*}
J(e(t_{k}),r(t_{k}),u(t_{k})) & = & \sum_{i=1}^{N}\left(e(t_{k+i})^{\top}Q_{e}e(t_{k+i})\right.\\
 &  & +r(t_{k+i})^{\top}Q_{r}r(t_{k+i})\\
 &  & \left.+u(t_{k+i})^{\top}Ru(t_{k+i})\right)
\end{eqnarray*}
subjected to the model dynamics discretized using Euler's method with
a step size of 0.01 seconds. The controller was implemented using
MATLAB's fmincon optimizer with $Q_{e}=I$, $Q_{r}=I$, $R=0.0001I$
and a prediction horizon of $N=5$ steps and bounded control input
search space with upper and lower bounds of 50 and -50, respectively,
for every control input.

\subsection{UUV System}

The simulations were performed on an unmanned underwater vehicle (UUV)
system that can be modeled as \cite{Fischer.Hughes.ea2014} 
\begin{eqnarray}
\ddot{x} & = & -\overline{M}^{-1}\left(x\right)\left(\overline{C}\left(x,\dot{x},\nu\right)\dot{x}+\overline{D}\left(x,\nu\right)\dot{x}+\overline{G}\left(x\right)\right)\nonumber \\
 &  & +\overline{M}^{-1}\left(x\right)\tau_{n},\label{eq:UUV earth fixed}
\end{eqnarray}
where $x\in\rr^{6}$ denotes a vector of position and orientation
with coordinates in the earth-fixed frame, $\dot{x}\in\rr^{6}$ denotes
a vector of linear and angular velocities with coordinates in the
earth-fixed frame, and $\nu\in\rr^{6}$ denotes a vector of linear
and angular velocities with coordinates in the body-fixed frame. The
inertial effects, centripetal-Coriolis effects, hydrodynamic damping
effects, gravitational effects, and control input in the earth-fixed
frame can be represented by $\overline{M}:\rr^{6}\rightarrow\rr^{6\times6}$,
$\overline{C}:\rr^{6}\times\rr^{6}\times\rr^{6}\rightarrow\rr^{6\times6}$,
$\overline{D}:\rr^{6}\times\rr^{6}\rightarrow\rr^{6\times6}$, $\overline{G}:\rr^{6}\to\rr^{6}$,
and $\tau_{n}:\rr_{\geq0}\rightarrow\rr^{6}$, respectively. The velocities
in the body-fixed frame can be related to the velocities in the earth-fixed
frame using the relation
\begin{align}
\dot{x} & =J\left(x\right)\nu,\label{eq:body to earth frame}
\end{align}
where $J:\rr^{6}\to\rr^{6\times6}$ is a Jacobian transformation matrix
relating the two frames \cite[Equation (2)]{Fischer.Hughes.ea2014}.
Thus, the dynamics in (\ref{eq:UUV earth fixed}) can be represented
using Eq. (1) from the manuscript with 
\[
f(x,\dot{x})=-\overline{M}^{-1}\left(x\right)\left(\overline{C}\left(x,\dot{x},\nu\right)\dot{x}+\overline{D}\left(x,\nu\right)\dot{x}+\overline{G}\left(x\right)\right)
\]
and 
\[
g(x,\dot{x})=\overline{M}^{-1}\left(x\right).
\]
Using the kinematic transformation in (\ref{eq:body to earth frame}),
the earth--fixed dynamics in (\ref{eq:UUV earth fixed}) can be expressed
using body-fixed dynamics as $\overline{M}=J^{-\top}MJ^{-1},$ $\overline{C}=J^{-\top}\left[C\left(\nu\right)-MJ^{-1}\dot{J}\right]J^{-1}$,
$\overline{D}=J^{-\top}D\left(\nu\right)J^{-1}$, $\overline{G}=J^{-\top}G$,
and $\tau_{n}=J^{-\top}\tau_{b}$, where $M\in\rr^{6\times6}$, $C:\rr^{6}\rightarrow\rr^{6\times6}$,
$D:\rr^{6}\rightarrow\rr^{6\times6}$, $G:\mathbb{R}^{6}\to\mathbb{R}^{6}$,
and $\tau_{b}:\rr_{\geq0}\rightarrow\rr^{6}$ denote the inertial
effects, centripetal-Coriolis effects, hydrodynamic damping effects,
gravitational effects, and control input in the body-fixed frame,
respectively. The inertial effects, centripetal-Coriolis effects,
and hydrodynamic damping effects in the body-fixed effects can be
expressed as \cite[Equation (2.246)]{Dixon2003}{\small{} 
\begin{align*}
M & =\text{diag}\left\{ m_{1},m_{2},m_{3},m_{4},m_{5},m_{6}\right\} ,\\
D & =\text{diag}\left\{ d_{11}+d_{12}\left|\nu\left(1\right)\right|,d_{21}+d_{22}\left|\nu\left(2\right)\right|,d_{31}+d_{32}\left|\nu\left(3\right)\right|,\right.\\
 & \quad\quad\left.d_{41}+d_{42}\left|\nu\left(4\right)\right|,d_{51}+d_{52}\left|\nu\left(5\right)\right|,d_{61}+d_{62}\left|\nu\left(6\right)\right|\right\} ,
\end{align*}
}{\scriptsize
\[
V_{m}=\left[\begin{array}{cccccc}
0 & 0 & 0 & 0 & m_{3}\nu_{3} & -m_{2}\nu_{2}\\
0 & 0 & 0 & -m_{3}\nu_{3} & 0 & m_{1}\nu_{1}\\
0 & 0 & 0 & m_{2}\nu_{2} & -m_{1}\nu_{1} & 0\\
0 & m_{3}\nu_{3} & -m_{2}\nu_{2} & 0 & m_{6}\nu_{6} & -m_{5}\nu_{5}\\
-m_{3}\nu_{3} & 0 & m_{1}\nu_{1} & -m_{6}\nu_{6} & 0 & m_{4}\nu_{4}\\
m_{2}\nu_{2} & -m_{1}\nu_{1} & 0 & m_{5}\nu_{5} & -m_{4}\nu_{4} & 0
\end{array}\right],
\]
}where the numerical values of mass, inertia, and damping parameters
listed in Table \ref{tab:UUV-System-Parameters} were used. The considered
UUV is neutrally buoyant, thus $G=0_{6\times1}$. The desired trajectory
was selected as a helical trajectory given by $x_{d}(t)=[2\text{cos}\left(0.5t\right)\text{m},2\text{sin\ensuremath{\left(0.5t\right)\text{m}},}0.1t\:\text{m},0\:\text{rad},0\text{\:rad},-0.125t\text{\:rad}]^{\top}$,
and the system was initialized with $x(0)=[-0.5\,\mathrm{m},-0.5\,\mathrm{m},-0.5\,\mathrm{m},0\,\mathrm{rad},0\,\mathrm{rad},0\,\mathrm{rad}]^{\top}$
and $\dot{x}(0)=[0_{1\times3}\,\mathrm{m/s},0_{1\times3}\,\mathrm{rad/s}]^{\top}$.
The following gains were used in the simulation: $\alpha_{1}=5$,
$\alpha_{2}=10$, $\alpha_{3}=40$, $k_{r}=20$, $k_{f}=20$, $k_{\hat{\theta}}=0.0001$,
$\Gamma(0)=0.5I_{221}$, $\beta=10$. The weights are randomly initialized
from the distribution $U(-0.5,0.5)$. Similar to the two-link manipulator,
for a fair comparison, the set of gains common to the developed and
baseline methods were selected to be exactly the same. The MPC was
implemented in a similar manner as the two-link manipulator (see Appendix
IX.1.1 for details), except with optimizer with $Q_{e}=I$, $Q_{r}=I$,
$R=10I$, $N=5$ steps, and bounded control input search space with
upper and lower bounds of 5 N and -5 N, respectively, for every linear
control input, and 5 Nm and -5 Nm, respectively, for every angular
control input, as these values were empirically found to yield the
most desirable performance.
\begin{table}
\centering{}\caption{\label{tab:UUV-System-Parameters}UUV System Parameters \cite[Equation (2.247)]{Dixon2003}}
\begin{tabular}{|c|c|c|}
\hline 
$m_{1}=215$ kg & $d_{11}=70$ Nm$\cdot$sec & $d_{41}=30$ Nm$\cdot$sec\tabularnewline
\hline 
$m_{2}=265$ kg & $d_{12}=100$ N$\cdot$sec$^{2}$ & $d_{42}=50$ N$\cdot$sec$^{2}$\tabularnewline
\hline 
$m_{3}=265$ kg & $d_{21}=100$ Nm$\cdot$sec & $d_{51}=50$ Nm$\cdot$sec\tabularnewline
\hline 
$m_{4}=40$ kg$\cdot\text{m}^{2}$ & $d_{22}=200$ N$\cdot$sec$^{2}$ & $d_{52}=100$ N$\cdot$sec$^{2}$\tabularnewline
\hline 
$m_{5}=80$ kg$\cdot\text{m}^{2}$ & $d_{31}=200$ Nm$\cdot$sec & $d_{61}=50$ Nm$\cdot$sec\tabularnewline
\hline 
$m_{6}=80$ kg$\cdot\text{m}^{2}$ & $d_{32}=50$ N$\cdot$sec$^{2}$ & $d_{62}=100$ N$\cdot$sec$^{2}$.\tabularnewline
\hline 
\end{tabular}
\end{table}

\section{Dwell-Time Conditions for Stable Operation under Intermittent Loss
of State Feedback}

Since the DNN identifies the system dynamics, the identified DNN estimates
could be used to predict the uncertainty when the state feedback is
intermittently lost. Let $i\in\mathbb{Z}_{\geq0}$ denote the time
index such that the state feedback is available in the time interval
$[t_{2i},t_{2i+1})$ and unavailable in the time interval $[t_{2i+1},t_{2i+2})$
for all $i\in\mathbb{Z}_{\geq0}$. During the time interval $[t_{2i},t_{2i+1})$,
when the feedback is available, the developed composite adaptive control
and adaptation laws developed in the manuscript are used for all $i\in\mathbb{Z}_{\geq0}$.
However, during the time interval $[t_{2i+1},t_{2i+2})$ when the
state feedback is unavailable, the control input is designed to be
an open-loop controller based on the last DNN weight estimate that
was identified the feedback was available. The open-loop controller
is given by 
\begin{eqnarray}
u & = & g^{+}\left(x_{d}{(t)},\dot{x}_{d}{(t)}\right)\left(\ddot{x}_{d}{(t)}-\Phi\left(X_{d}{(t)},\hat{\theta}(t_{2i+1})\right)\right).\label{eq:open loop control}
\end{eqnarray}

Substituting (\ref{eq:open loop control}) into $\ddot{x}=f(x,\dot{x})+g(x,\dot{x})u$,
subtracting $\ddot{x}_{d}$ on both sides, adding and subtracting
$\Phi\left(X_{d},\hat{\theta}(t_{2i+1})\right)$, and rearranging
terms yields 
\begin{eqnarray}
\ddot{e} & = & \Phi(X,\theta^{*})-\Phi\left(X_{d}{(t)},\hat{\theta}(t_{2i+1})\right)+\varepsilon(X)\nonumber \\
 &  & +\left(g(x,\dot{x})g^{+}\left(x_{d}{(t)},\dot{x}_{d}{(t)}\right)-I_{n}\right)\left(\ddot{x}_{d}{(t)}\right.\\
 &  & -\Phi\left(X_{d}{(t)},\hat{\theta}(t_{2i+1})\right).\label{eq:error system without feedback}
\end{eqnarray}
For the purpose of this section, it is assumed the drift dynamics
$f$ are globally Lipschitz with a Lipschitz constant $\varpi\in\mathbb{R}_{>0}$,
and the control effectiveness and its pseudoinverse, $g$ and $g^{+}$,
are globally bounded functions without bounds $\overline{g},\overline{g^{+}}$
such that $\left\Vert g(x,\dot{x})\right\Vert \leq\overline{g}$ and
$\left\Vert g^{+}(x,\dot{x})\right\Vert \leq\overline{g^{+}}$. The
global Lipschitzness of $f$ is assumed in order to rule out the possibility
of the drift dynamics causing finite-time escape during the absence
of state-feedback. Such an assumption is reasonable since finite-time
escape is usually not inherent to the uncontrolled dynamics for most
practical systems of interest. Additionally, assuming that $g$ and
$g^{+}$ are bounded is reasonable for most practical engineering
systems, since the control effectiveness term usually results from
the inertia matrix or the kinematic Jacobian of the system, and systems
that may have potentially singular kinematic Jacobians in practice
are not considered here. Additionally, in this section, a requirement
is imposed on the selected DNN $\Phi$ to contain bounded globally
Lipschitz activation functions. Thus, using bounds on $g$, $g^{+}$,
$\Phi\left(X_{d},\hat{\theta}(t_{2i+1})\right)$, $\tilde{\theta}(t_{2i+1})$,
and $\ddot{x}_{d}$, it can be shown that there exists constants $L_{U},\delta_{U}\in\mathbb{R}_{>0}$
such that $\left\Vert \ddot{e}\right\Vert \leq L_{U}\left\Vert e\right\Vert +L_{U}\left\Vert \dot{e}\right\Vert +\delta_{U}.$
Using the relations $r=\dot{e}+\alpha_{1}e$ and $\ddot{e}=\dot{r}-\alpha_{1}\dot{e}$
yields the inequality $\left\Vert \dot{r}\right\Vert \leq\left(\alpha_{1}^{2}+L_{U}\alpha_{1}+L_{U}\right)\left\Vert e\right\Vert +\left(L_{U}+\alpha_{1}\right)\left\Vert r\right\Vert +\delta_{U}.$
Additionally, since $f$ is considered to be globally Lipschitz in
this section, it follows that $\left\Vert \frac{\partial f}{\partial x}\right\Vert ,\left\Vert \frac{\partial f}{\partial\dot{x}}\right\Vert \leq\varpi$.
As a result, it can be shown that $\ensuremath{\left\Vert \dot{f}\right\Vert \leq\left(2\alpha_{1}^{2}+\left(L_{U}+1\right)\alpha_{1}+L_{U}\right)\varpi\left\Vert e\right\Vert +\left(L_{U}+2\alpha_{1}+1\right)\varpi\left\Vert r\right\Vert +\left(\delta_{U}+\overline{\dot{x}_{d}}+\overline{\ddot{x}_{d}}\right)\varpi}$.
During the loss of state feedback, all observer and adaptive update
laws are selected to be zero, i.e., $\dot{\ensuremath{\hat{r}}}=0$,
$\dot{\ensuremath{\hat{f}}}=0$, and $\dot{\ensuremath{\hat{\theta}}}=0$.

The growth of the Lyapunov function in (\ref{eq:Lyap}) is examined
using the bounds on $\dot{r}$ and $\dot{f}$ to analyze the growth
of the error states during the loss of state feedback. By successive
use of Holder's and Young's inequalities, it can be shown that $\dot{V}\leq\lambda_{U}V+\Delta_{U},$
when feedback is unavailable, where $\lambda_{U}\triangleq2\max(\frac{3\alpha_{1}^{2}+L_{U}\alpha_{1}+L_{U}+1}{2}+(2\alpha_{1}^{2}+\left(L_{U}+1\right)\alpha_{1}+L_{U})^{2}\varpi^{2}+(\alpha_{1}^{2}+L_{U}\alpha_{1}+L_{U})^{2},\frac{\alpha_{1}^{2}+\left(L_{U}+2\right)\alpha_{1}+3L_{U}+\delta_{U}+1}{2}+(L_{U}+2\alpha_{1}+1)^{2}\varpi^{2}+(L_{U}+\alpha_{1})^{2},\frac{1}{2})$
and $\Delta_{U}\triangleq\frac{\delta_{U}}{2}+\delta_{U}^{2}+(\delta_{U}+\overline{\dot{x}_{d}}+\overline{\ddot{x}_{d}})^{2}\varpi^{2}$.
Solving for $V$ for yields $V(t)\leq V(t_{2i+1})\mathrm{e}^{\lambda_{U}(t-t_{2i+1})}+\frac{\Delta_{U}}{\lambda_{U}}(\mathrm{e}^{\lambda_{U}(t-t_{2i+1})}-1)$
for all $(t,i)\in[t_{2i+1},t_{2i+2})\times\mathbb{Z}_{\geq0}$. Then
applying the bounds in (\ref{eq:Lyap bounds}) and taking the square
root yields {\footnotesize
\begin{equation}
\left\Vert z(t)\right\Vert \leq\sqrt{\frac{\lambda_{2}}{\lambda_{1}}\left\Vert z(t_{2i+1})\right\Vert ^{2}\mathrm{e}^{\lambda_{U}\left(t-t_{2i+1}\right)}+\frac{2\Delta_{U}}{\lambda_{1}\lambda_{U}}\left(\mathrm{e}^{\lambda_{U}\left(t-t_{2i+1}\right)}-1\right)},\label{eq:z unstable bounds}
\end{equation}
}for all $(t,i)\in[t_{2i+1},t_{2i+2})\times\mathbb{Z}_{\geq0}$.

When the system regains feedback, the condition $z(t_{2i+2})\in\mathcal{S}$
needs to be satisfied for the composite adaptive Lb-DNN to yield the
results in Theorem 1 of the manuscript. Imposing this condition yields
the following condition for maximum dwell time during which feedback
can be unavailable without affecting the UUB properties of the resulting
switched system, 
\begin{equation}
\left(t_{2i+2}-t_{2i+1}\right)\leq\frac{1}{\lambda_{U}}\ln\left(\frac{\frac{\lambda_{1}}{\lambda_{2}}\chi^{2}+\frac{2\Delta_{U}}{\lambda_{1}\lambda_{U}}-\frac{c}{\lambda_{3}}}{\frac{\lambda_{2}}{\lambda_{1}}\left\Vert z(t_{2i+1})\right\Vert ^{2}+\frac{2\Delta_{U}}{\lambda_{1}\lambda_{U}}}\right),\label{eq:max dwell time}
\end{equation}
for $(t,i)\in[t_{2i+1},t_{2i+2})\times\mathbb{Z}_{\geq0}$. The maximum
dwell time in (\ref{eq:max dwell time}) should be positive for the
system to sufficiently allow feedback unavailability, which holds
when $\frac{\lambda_{1}}{\lambda_{2}}\chi^{2}+\frac{2\Delta_{U}}{\lambda_{1}\lambda_{U}}-\frac{c}{\lambda_{3}}>\frac{\lambda_{2}}{\lambda_{1}}\left\Vert z(t_{2i+1})\right\Vert ^{2}+\frac{2\Delta_{U}}{\lambda_{1}\lambda_{U}}$.
Imposing this condition on $\left\Vert z(t_{2i+1})\right\Vert ^{2}$
and using Theorem 1 of the manuscript yields the following condition
for minimum dwell time during which the feedback should be available
\begin{equation}
(t_{2i+1}-t_{2i})\geq\frac{\lambda_{2}}{\lambda_{3}}\ln\left(\frac{\frac{\lambda_{2}}{\lambda_{1}}\left\Vert z(t_{2i})\right\Vert ^{2}}{\frac{\lambda_{1}^{2}}{\lambda_{2}^{2}}\chi^{2}-\frac{\lambda_{2}c}{\lambda_{1}\lambda_{3}}-\frac{\lambda_{1}c}{\lambda_{2}\lambda_{3}}}\right),\label{eq:min dwell time}
\end{equation}
for all $(t,i)\in[t_{2i+1},t_{2i+2})\times\mathbb{Z}_{\geq0}$. Note
that it is permissible to obtain negative values for the dwell-time
in (\ref{eq:min dwell time}), since a negative minimum dwell-time
for feedback availability would imply the stability guarantees hold
even if feedback continues to be unavailable after the time instance
$t_{2i}$. Additionally, the size of set $\mathcal{D}$, i.e., $\chi$
needs to be selected according to $\chi>\sqrt{\frac{\lambda_{2}^{3}c}{\lambda_{1}^{3}\lambda_{3}}+\frac{\lambda_{2}c}{\lambda_{1}\lambda_{3}}}$,
to ensure a positive denominator in (\ref{eq:min dwell time}), thus
guaranteeing the feasibility of the minimum dwell-time condition.
This results in the additional gain condition $k_{\min}>\lambda_{3}+\bar{\rho}\left(\sqrt{\frac{\lambda_{2}^{3}c}{\lambda_{1}^{3}\lambda_{3}}+\frac{\lambda_{2}c}{\lambda_{1}\lambda_{3}}}\right)$.
\begin{IEEEbiography}[{\includegraphics[width=1\columnwidth]{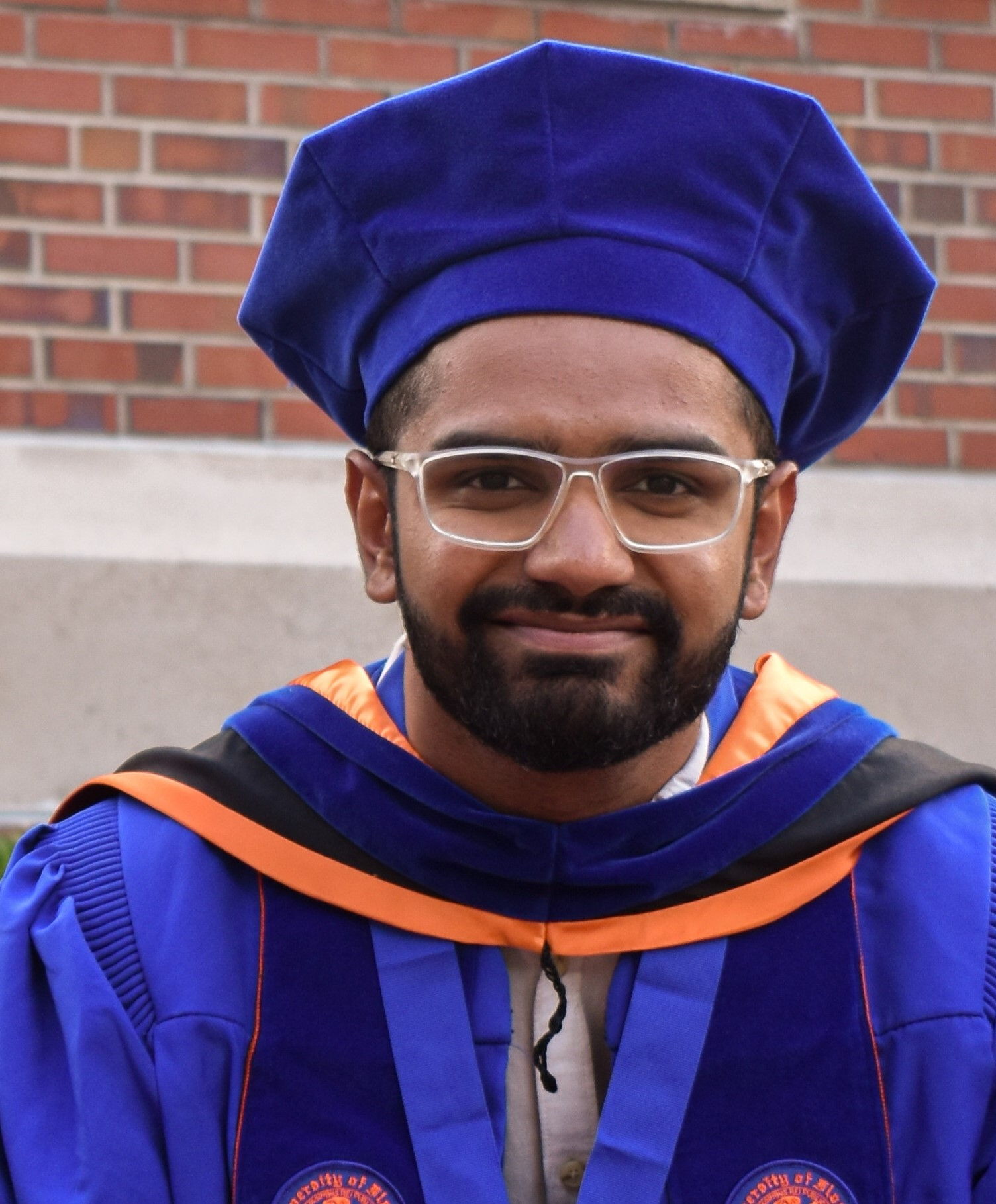}}]{Omkar Sudhir Patil}
 received his Bachelor of Technology (B.Tech.) degree in production
and industrial engineering from Indian Institute of Technology (IIT)
Delhi in 2018, where he was honored with the BOSS award for his outstanding
bachelor's thesis project. In 2019, he joined the Nonlinear Controls
and Robotics (NCR) Laboratory at the University of Florida under the
guidance of Dr. Warren Dixon to pursue his doctoral studies. Omkar
received his Master of Science (M.S.) degree in mechanical engineering
in 2022 and Ph.D. in mechanical engineering in 2023 from the University
of Florida. During his Ph.D. studies, he was awarded the Graduate
Student Research Award for outstanding research. In 2023, he started
working as a postdoctoral research associate at NCR Laboratory, University
of Florida. His research focuses on the development and application
of innovative Lyapunov-based nonlinear, robust, and adaptive control
techniques. 
\end{IEEEbiography}

\begin{IEEEbiography}[{\includegraphics[width=1\columnwidth]{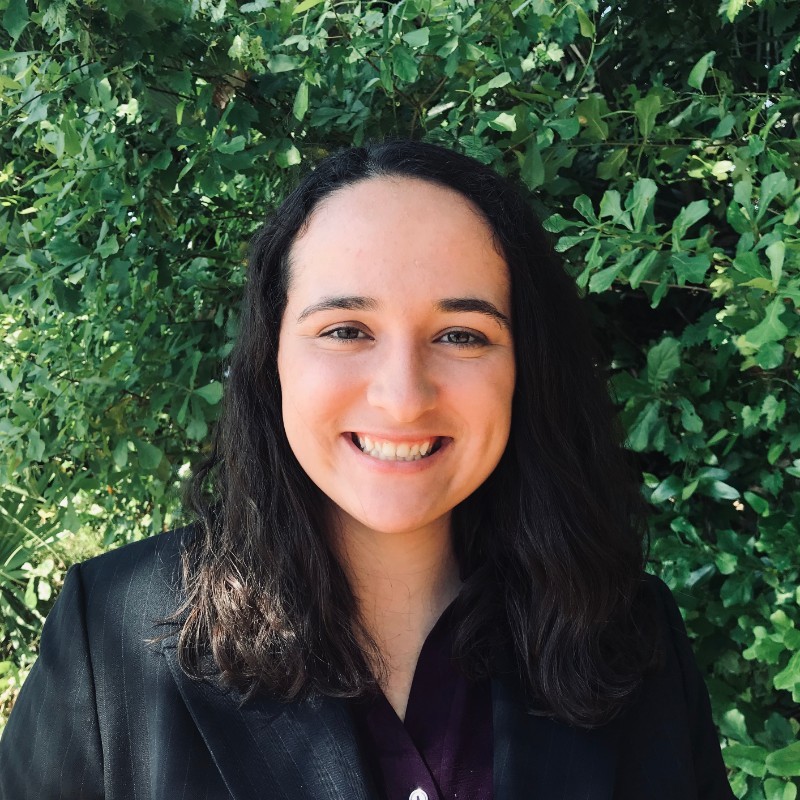}}]{Emily J. Griffis}
 received the B.S. and M.S. degrees in mechanical engineering and
the Ph.D. degree from the University of Florida, in May 2020, December
2021, and May 2024, respectively. In 2020, she joined the Nonlinear
Controls and Robotics Laboratory, University of Florida, under the
supervision of Dr. Warren Dixon to pursue her Ph.D. degree. Her research
interests include using adaptive control and deep learning to study
Lyapunov-based control of nonlinear and uncertain systems. In 2024,
she was awarded the Best Dissertation Award for the Department of
Mechanical and Aerospace Engineering.
\end{IEEEbiography}

\begin{IEEEbiography}[{\includegraphics[width=1\columnwidth]{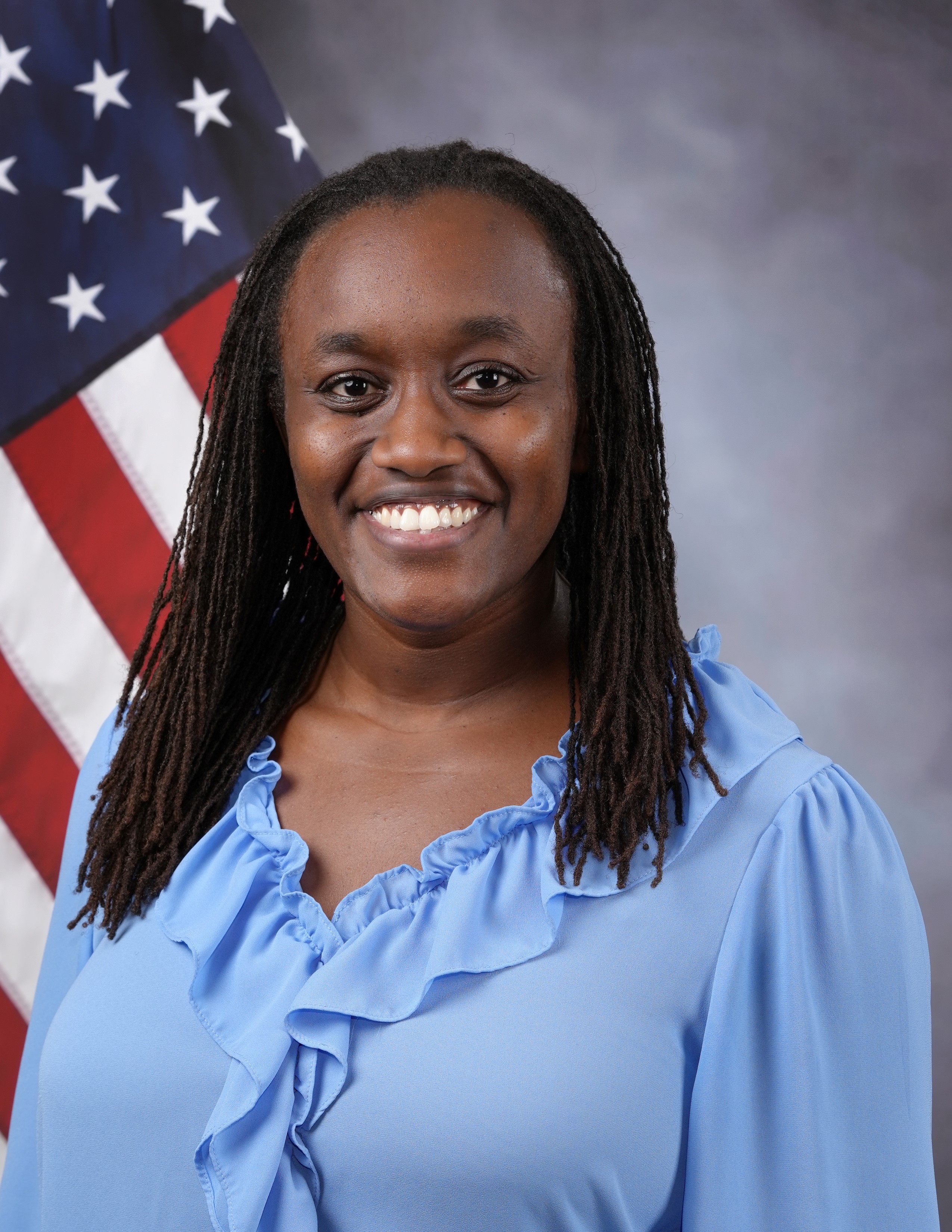}}]{Wanjiku A. Makumi}
 received her B.S. from the Joint Department of Biomedical Engineering
at NC State University and UNC Chapel Hill. In 2020, she joined the
Nonlinear Controls and Robotics Laboratory at the University of Florida
under the guidance of Dr. Warren Dixon where she received her M.S.
in mechanical engineering and her Ph.D. in aerospace engineering in
2021 and 2024, respectively. In 2025 she was awarded the Best Dissertation
Award for the Department of Mechanical and Aerospace Engineering.
She is a recipient of the DoD SMART Scholarship and is currently a
research engineer at the Air Force Research Laboratory. Her research
focuses on machine learning and adaptive control for unknown nonlinear
systems.
\end{IEEEbiography}

\begin{IEEEbiography}[{\includegraphics[width=1\columnwidth]{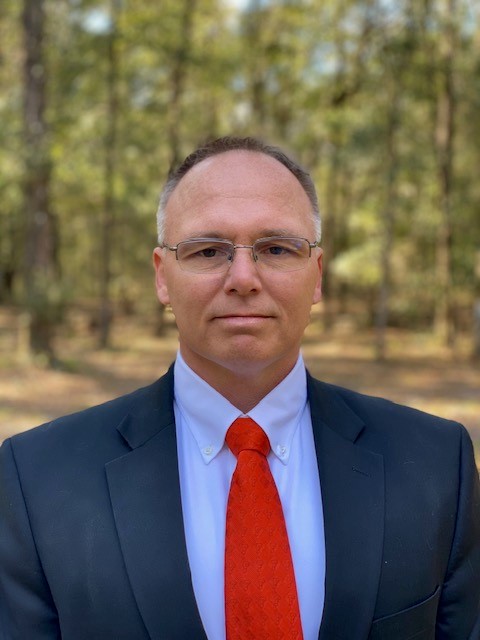}}]{Prof. Warren E. Dixon}
 received his Ph.D. from the Department of Electrical and Computer
Engineering from Clemson University. He worked as a research staff
member and Eugene P. Wigner Fellow at Oak Ridge National Laboratory
(ORNL) for four years until he joined the University of Florida in
the Mechanical and Aerospace Engineering Department. His main research
interest has been the development and application of Lyapunov-based
control techniques for uncertain nonlinear systems. He is an ASME
and IEEE Fellow and his collaborative work has received various early
and mid-career awards, including various best paper awards.
\end{IEEEbiography}

\end{document}